\documentclass[12pt]{iopart} 

\usepackage[utf8]{inputenc} 

\usepackage{graphicx}
\usepackage{subfigure}
\usepackage{iopams,amsthm, amssymb, amsfonts}

\pdfoutput=1
\usepackage{etex}
\reserveinserts{18}

\usepackage{cite}
\usepackage{latexsym}
\usepackage{rawfonts}
\usepackage[usenames,dvipsnames,svgnames]{xcolor}
\usepackage{bbm}
\usepackage{latexsym}
\usepackage{multirow}
\usepackage{rotating}
\usepackage{lscape}
\usepackage{fancybox}
\usepackage{ifmtarg}

\usepackage{rotating}

\usepackage{cmbright}
\usepackage{textcomp}
\usepackage[font=small,labelfont=sf]{caption}

\def\norf{\normalfont}

\def\edge#1#2{{\langle #1{\sim}#2 \rangle}}
\def\sfrac#1#2{\hbox{\normalsize $\frac{#1}{#2}$}}
\def\Sfrac#1#2{\hbox{\large$\frac{#1}{#2}$}}
\def\SSfrac#1#2{\hbox{\Large$\frac{#1}{#2}$}}

\def\Ref#1{(\ref{#1})}
\def\thin{\:\!}

\begin{document}
\title{Mean Field Analysis of Algorithms for Scale-free Networks in Molecular Biology}
\author{S Konini and EJ Janse van Rensburg\footnote[1]{\texttt{rensburg@yorku.ca}}}
\address{\norf Department of Mathematics and Statistics, 
York University, Toronto, Ontario M3J~1P3, Canada\\}

\begin{abstract} 
\norf
The sampling of scale-free networks in Molecular Biology is usually achieved by
growing networks from a seed using recursive algorithms with elementary moves
which include the addition and deletion of nodes and bonds.  These algorithms
include the Barabasi-Albert algorithm.  Later algorithms, such as the
Duplication-Divergence algorithm, the Sol\'e algorithm and the iSite algorithm,
were inspired by biological processes underlying the evolution of protein networks,
and the networks they produce differ essentially from networks grown by the
Barabasi-Albert algorithm.  In this paper the mean field analysis of these algorithms
is reconsidered, and extended to variant and modified implementations of the
algorithms.  The degree sequences of scale-free networks decay according to
a powerlaw distribution, namely $P(k) \sim k^{-\gamma}$, where $\gamma$ is
a scaling exponent.  We derive mean field expressions for $\gamma$,
and test these by numerical simulations.   Generally, good agreement is obtained.
We also found that some algorithms do not produce scale-free networks
(for example some variant Barabasi-Albert and Sol\'e networks).
\end{abstract}



\maketitle



\section{Introduction}

Many systems in nature and society are described by means of complex networks  
\cite{BLM06}. Some of these systems include the cell \cite{JTA00}, 
chemical reactions \cite{GA05}, the world wide web \cite{BAJ00}, 
social interactions \cite{BBP04}, etc.  It is generally found that many system,
though different in nature,  produce networks which are scale-free and
exhibit similar properties \cite{AB02,B09}. 

The main property of scale-free networks is that their degree distribution 
decays as a power law \cite{AB02,BA99} -- this shows that there is no
characteristic scale for the degrees, which is why the networks are called scale-free.
The average degree of a scale-free network offers little insight into the 
real topology of the network \cite{B09} since most nodes have degrees which are far 
away from the average degree of the network.  Nodes of high degree are
called \textit{hubs} and though small in number for realistic networks, they are
over-represented compared to the number of hubs in random networks.  These
hubs play an important role in dynamical processes which occur in scale-free networks.

Scale-free networks also exhibit an unexpected degree of robustness --  this is the
property that such networks maintain their dynamic properties even when many 
nodes and bonds fail to transmit signals (suffer high failure rates) \cite{BLM06}.
However, these networks remain vulnerable to failure of hub nodes, since these
nodes play a significant role in maintaining the network's connectivity.

In this paper the mean field approach to the analysis of algorithms for sampling
scale-free networks inspired by processes in molecular biology is presented.
In addition, numerical testing and, in some cases, verification, of the mean field
approach will be examined.  The focus will be on four widely used and discussed
algorithms in the literature, nameley, the Barabasi-Albert algorithm \cite{BA99,BAJ99}, 
the Duplication-Divergence algorithm \cite{TR04,VF02}, 
the Sol\'e algorithm \cite{SPSK02} and the iSite algorithm \cite{G11,GG15}. 

The Duplication-Divergence, Sol\'e and iSite algorithms are inspired by modelling
networks in biological models of protein-protein interaction evolution, and all these
algorithms are based in one way or another on two ideas: growth by 
preferential attachment \cite{EL03}, and growth and changes (\textit{mutations})
in networks induced by the duplication, deletion or replacement of nodes or bonds (these
are elementary moves which \textit{mutate} the network by adding, deleting 
or moving some of its bonds or nodes). 

Growth by preferential attachment is implemented by adding bonds preferentially to
nodes of high degree.  This increases the probability that a node will grow to be
a hub in the network, and the resulting network has an increased probability that
it will contain hubs \cite{BA99}.  The Barabasi-Albert algorithm uses preferential
attachment to grow scale-free networks by attaching bonds to nodes with a probability
which is proportional to the degrees of nodes \cite{AB02}.  A mean field analysis
of the Barabasi-Albert algorithm was done in reference \cite{BAJ99}.

The Duplication-Divergence algorithm \cite{VF02,TR04} generates scale-free networks
by implementing elementary moves which mutate and grow the network.  These
are \textit{duplication} (the duplication of existing nodes and bonds) and \textit{divergence}
(local changes made to existing bonds and nodes) elementary moves.  These moves model 
processes which are thought to underlie the evolutionary mechanisms by which protein
interaction networks evolve \cite{VF02,PS03,TR04}:  The \textit{duplication} of genes 
is a mechanism which generates genes coding for new proteins during evolution
and the \textit{divergence} step is a model for the mutation of duplicated genes. 
After a duplication of a gene, two genes (one the \textit{progenitor} gene, the other the
\textit{progeny} gene) coding for the same protein are obtained, and these
mutate over time to drift away from one another in gene space, giving rise to modified
proteins when translated by cellular machinery \cite{PS03}.  Biologically, the
duplication step may result in a new protein interaction between two mutating copies
of the same gene (this is called heteromerization), and the divergence
step is a model of subfunctionalization (a process where interactions between
proteins are lost).

Closely related to the Duplication-Divergence algorithm is the Sol\'e algorithm 
\cite{SPSK02,PS03}.  This algorithm grows networks by duplication of nodes, and 
mutates the network by rewiring it (this algorithm does not implement the heteromerization
of the duplicated genes)  \cite{BAJ00}.  It then implements a process of deleting some bonds
on the duplicated nodes (modelling evolutionary changes due to subfunctionalization).  

The iSite algorithm \cite{G11,GG15} is a refinement of the Duplication-Divergence and
Sol\'e algorithms.  This algorithm introduces more complex nodes which each contains
\textit{interaction sites} as models of protein and protein complexes with localized 
interaction sites where the interactions with other proteins take place.  These localized
interaction sites are \textit{iSites}.  Such iSites may be involved in many interactions, but
each interaction is related to only two iSites, one on each of the proteins involved.
That is, iSites are models of the concept of domains on protein surfaces where the
actual interactions take place between two proteins.  The implementation of the
algorithm on nodes containing iSites proceeds by duplication of nodes, and the mutation
of iSites through subfunctionalization and heteromerization
(namely, the subfunctionalization of iSites leading to loss of
protein interactions, and heteromerization where new interactions are introduced
between existing iSites).  In this model the subfunctionalization is of iSites, leading to
the loss of all bonds incident with the iSite (contrary to the situation in other algorithms,
for example the Duplication-Divergence algorithm, where subfunctionalization leads
to the loss of bonds, rather than nodes).

This paper is organised as follows.  We first consider the general properties of
scale-free networks, including their scaling and connectivity properties.  These ideas
are then applied to the analysis of particular algorithms.  The Barabasi-Albert
model is considered first together with a modified version of the algorithm, and a
variant of the algorithm.  Mean field theory for the modified and variant
algorithms is developed, giving mean field values for the scaling exponent
$\gamma$.  These results are compared to numerical results obtained by generating
networks using implementations of the algorithms.

The Duplication-Divergence algorithm and networks generated by it are considered
next.  The algorithm is also modified, and mean field theory is developed to
find mean field values for the scaling exponent.  The mean field predictions are
then compared to numerical results generated by implementing the algorithm
and sampling networks.

A similar approach is followed for the Sol\'e algorithm.  However, in this
model the degree distribution is not integrable, and our results indicate that
the networks generated by this algorithm are not scale-free.  Instead, the degree
distribution must be modified.  This gives a testable scaling hypothesis for 
Sol\'e networks, which is tested numerically by generating networks and
examining their scaling, as well as by computing the connectivity of Sol\'e
networks and comparing it to the mean field predictions.  This shows that
the size of Sol\'e networks of order $n$ is $O(n^2)$, while the connectivity
is $O(n)$ -- this implies that Sol\'e networks are rich in bonds.

Finally, the iSite algorithm is presented and examined developing a mean
field approach to determine its scaling properties.  The algorithm is also 
modified, and the resulting mean field results are tested numerically.

The paper is completed in the conclusion section, where our main results
are briefly considered and reviewed.

\section{Scale-free networks}

Scale-free networks of order $n$ are characterised by degree sequences 
$\{d_k \}$ which follow a power law distribution (where $d_k$ is the number of 
nodes of degree $k$ and $\sfrac{1}{n} d_k$ is the fraction of nodes of 
degree $k$).

If $\langle d_k \rangle$ is the average degree distribution, then 
$\sfrac{1}{n} \langle d_k \rangle$ is proportional to the probability 
$P(k)$ that a node has degree $k$. In scale-free networks, the
probability $P(k)$ decays like a powerlaw with exponent $\gamma$:
\begin{equation}
P(k) \simeq C_o^{-1}\, k^{-\gamma} .
\label{eqn1}  
\end{equation}
Here, $\gamma$ is the \textit{scale-free network exponent}.   The constant $C_o$ is a
normalisation constant given by
\begin{equation}
C_o = \sum_{k=1}^n P(k) .
\end{equation}
As $n\to\infty$, it is necessary that $\gamma > 1$ for $P(k)$ to be summable 
(and $C_o<\infty$).   In this case $C_o$ converges to a constant as $n\to\infty$.
Thus, if $\gamma>1$ then the network is said to be integrable 
with scaling exponent $\gamma$ (in this event equation \Ref{eqn1} is the scaling of 
the limiting degree distribution with $C_o>0$ finite and $P(k) \to 0$ as $k\to\infty$).

The case that $\gamma = 1$ gives rise to a logarithmic correction.  Since
$\sum_{k=1}^n k^{-1} \sim \log n$, this gives the distribution
\begin{equation}
P(k) \sim \Sfrac{1}{\log n}\, k^{-1}
\end{equation}
for networks of (large) order $n$.  This network is said to be not integrable, but
for asymptotic values and fixed values of $n$ the decay of $P(k)$ will appear 
to be proportional to $k^{-1}$.

Since $P(k)$ is the probability that a node in a network has degree $k$, the average
degree sequence $\{ \langle d_k \rangle_n \}$ over randomly generated networks of 
order $n$ is given approximately by $\langle d_k \rangle \sim n\thin P(k)$, 
for $n$ large.  It is not known that 
the degree sequence is self-averaging (that is, that the degree sequence $\{ d_k \}$ has 
distribution $d_k \sim n\thin P(k)$ as $n\to\infty$ for a single randomly generated
scale-free network).

This powerlaw decay of degree sequences shows that nodes of large degree 
(that is, for large $k$) are more common in scale-free networks (compared to 
randomly generated networks, where they are exponentially rare).  
These nodes of  large degree are called \textit{hubs}.  A precise definition of 
a hub in a network is somewhat arbitrary, but for the purpose of this paper, 
a ``hub" in a network of order $n$ is defined as a node of degree 
exceeding $\lfloor \sqrt{n} \rfloor$.

The exponent $\gamma$ can be estimated from numerical data by computing
the average degree sequence $\{ \langle d_k \rangle\}$ and then plotting 
$\log P(k) / \log k$ against $1/\log k$ (for networks of order $n\gg k$).  
Extrapolating the data to $k=\infty$ 
using a linear or a quadratic regression gives the value of $\gamma$ as the
$y$-intercept of the graph.  This method works well if $P(k)$ scales with $k$ as in
equation \Ref{eqn1}.  However, strong corrections to the powerlaw
behaviour may make the extrapolation difficult or inaccurate.  

A second method to estimate $\gamma$ is to note that if $\gamma>1$, then
for a fixed value of $\alpha>0$,
\begin{equation}
\zeta(k) = \log P(\alpha \, k) - \log P(k) = -\gamma \log \alpha + o(1) .
\label{eqn2}   
\end{equation}
Experimentation with numerical data shows that by plotting $\zeta(k)$ 
against $\sfrac{1}{k}\log k$ good results are obtained, and linear or
quadratic regressions of $\zeta(k)$ against $\sfrac{1}{k} \log k$ can be used to 
estimate $\gamma$.

If it is assumed that $P(k)$ is well approximated by equation \Ref{eqn1} for all 
$k\geq 1$, then the average \textit{connectivity} of a network of order $n$ with 
average degree distribution proportional to $P(k) = C_o n^{-\gamma}$ is given by
\begin{eqnarray}
\label{eqn3}   
\langle k \rangle_n &=
\SSfrac{\sum_{k=1}^n k\, P(k)}{\sum_{k=1}^n  P(k)}
\simeq \SSfrac{\int_1^n k\, P(k)\, dk}{\int_1^n  P(k)\, dk}
\simeq \left(\Sfrac{\gamma-1}{\gamma-2}\right)  \Sfrac{n^\gamma-n^2}{n^\gamma-n}
\nonumber \\
& \simeq \cases{
\left(\Sfrac{\gamma-1}{2-\gamma}\right) n^{2-\gamma}, & {if $1<\gamma<2$};\\
\left(\Sfrac{\gamma-1}{\gamma-2}\right) , & {if $\gamma>2$} .
} 
\end{eqnarray}
Observe that the asymptotic estimate is very poor if $\gamma \approx 2$, and if $n$
is small.

The cases $\gamma=1$ and $\gamma=2$ can also be determined; this gives
\begin{equation}
\langle k \rangle_n \simeq
\cases{
\Sfrac{n}{\log n}, & \hbox{if $\gamma=1$}; \\
\log n,  & \hbox{if $\gamma=2$}.
}
\label{eqn4}   
\end{equation}
The coefficient $\sfrac{\gamma-1}{\gamma-2}$ may be modifed if $P(k)$ is not well 
approximated by the powerlaw decay for smaller values of $k$ in equation \Ref{eqn1}.  
These results, however, do show that the connectivity is a constant independent of $n$ 
(for large $n$) if $\gamma>2$.

The expected number of bonds in the network is given by $E_n 
= \sfrac{1}{2}n\thin \langle k \rangle_n$.  Assuming the powerlaw relation in equation
\Ref{eqn1}, it follows that
\begin{eqnarray}
E_n =
\cases{
\Sfrac{n^2}{2\log n}, & \hbox{if $\gamma=1$}; \\
\left(\Sfrac{\gamma-1}{2(\gamma-2)}\right) n^{3-\gamma}, & \hbox{if $1<\gamma<2$};\\
\frac{1}{2} n\log n ,  & \hbox{if $\gamma=2$}; \\
\left(\Sfrac{\gamma-1}{2(\gamma-2)}\right) n , & \hbox{if $\gamma>2$} .
}
\label{eqn37}   
\end{eqnarray}
Of course, if $\gamma<1$, then $E_n = \Theta(n^2)$ and since a complete graph has
$\sfrac{1}{2}n(n-1)$ bonds, this implies that these graphs are dense in the sense that
$\liminf_{n\to\infty} \sfrac{1}{n^2} E_n >0$.  For all values of $\gamma\geq 1$ the
above shows that $\limsup_{n\to\infty} \sfrac{1}{n^2} E_n =0$, and the
graphs are sparse.

These results are useful in examining numerical data for scale-free networks.  For example,
$\gamma$ can be estimated by examining degree sequences averaged over randomly
sampled networks (from equation \Ref{eqn1}), or alternatively by using equation \Ref{eqn2}.
The connectivity $\langle k \rangle_n$ approaches a constant if $\gamma>2$ (as 
in equation \Ref{eqn3}) or grows as a powerlaw with $n$ if $\gamma<2$, and with
logarithmic corrections if $\gamma=1$ or $\gamma=2$ (as in equation \Ref{eqn4}).
Alternatively, the average size $E_n$ (the number of bonds in a network of order $n$)
can be considered, using the results in equation \Ref{eqn37}.

\section{Barabasi-Albert networks and the Barabasi-Albert algorithm}

The Barabasi-Albert algorithm is a recursive algorithm which grows networks (or
clusters of nodes and bonds) from a seed node.  This algorithm was introduced
in reference \cite{BA99} and reviewed in 2002 in a seminal paper \cite{AB02},
and its elementary move was inspired by processes underlying the (presumed) evolution
of scale-free networks seen in the physical world.  The elementary move is
a preferential attachment of new nodes (and bonds) to hubs (nodes of
high degree) in the network.  The algorithm is initiated by a single node,
and then new nodes and bonds are recursively attached, with new bonds preferentially
attached to existing nodes of large degree. 

A Barabasi-Albert network of order $N$ nodes is grown as follows:

\vspace{1mm}
\noindent{\bf Barabasi-Albert algorithm:}
\begin{enumerate}
\item[\bf 1.] Initiate the network with one node $x_0$;
\item[\bf 2.] Suppose that the network consists of nodes $\{x_0,x_1,\ldots,x_{n-1}\}$
of degrees $\{k_0,k_1,\ldots,k_{n-1}\}$;
\item[\bf 3.] Append a new node $x_n$ by executing step (a) or step (b):
\begin{enumerate}
\item[\bf (a)] With probability $p$: Select $x_j$ uniformly and attach $x_n$ to it by
inserting the bond $\edge{x_j}{x_n}$;
\item[\bf (b)] With default probability $1- p$: Attach $x_n$ by adding bonds
$\edge{x_j}{x_n}$ independently with probability $\sfrac{k_j}{\sum_j k_j}$;
\end{enumerate} 
\item[\bf 4.] Repeat step 3 until a network of order $N$ is grown.
\qed
\end{enumerate}
\vspace{5mm}

Step 3(a) is a random attachment of a node and bond, and step 3(b) attaches
a node with bonds \textit{preferentially} to existing nodes of high degree.
The algorithm has a single parameter $p$.  If $p=1$ then the algorithm grows
acyclic (and connected) networks of order $N$ (these are random trees).

\begin{figure}[t!]
 \centering
\includegraphics[height=7cm]{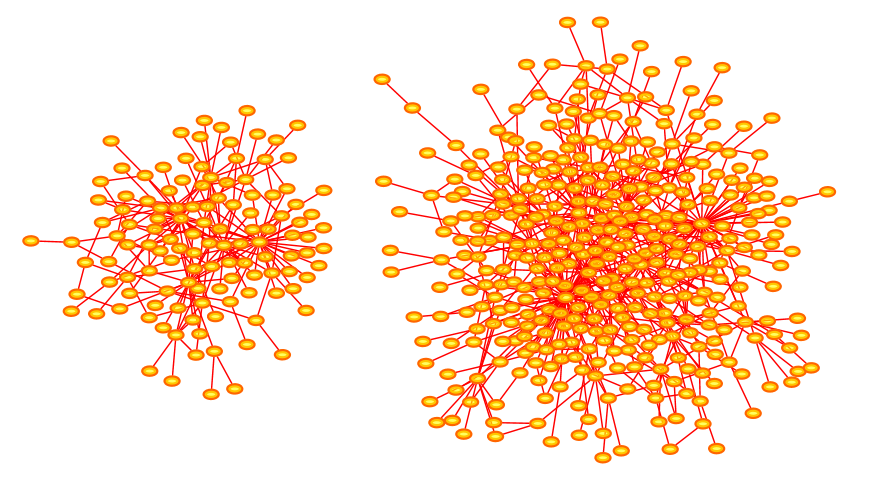}
\caption{{Barabasi-Albert networks with $p=0$:}
The network on the left was grown to order $n=122$.  It has 5 hubs of degrees $\{12,17,18,19,31\}$
exceeding $\sqrt{122}$.  The network on the right was grown to order $n=380$.  This
network has 3 hubs of degrees $\{29,47,63 \}$ exceeding $\sqrt{380}$.  
The arrangement of nodes and bonds in these networks was created using 
the prefuse force directed lay-out in Cytoscape 3.4.0
\cite{Cytoscape}.}
\label{figure1}
\end{figure}

On the other hand, if $p=0$, then step 3(b) is executed on each iteration.
New bonds are created with probabilities $q_j=\sfrac{k_j}{\sum_j k_j}$ for
$j=0,1,\ldots,n-1$ when the $n$-th node is added.  This shows that 
the expected number of bonds added in this step is on average
$\sum_j q_j = 1$.  That is, on average $1$ bond is added in each iteration, and the average
sum of degrees $\sum_j k_j$ should be equal to $2n$ by handshaking after $n$ iterations.  
This suggests that the algorithm grows a sparse graph with increasing $n$.  
However, since bonds are appended preferentially on growing hubs, 
the largest clusters in the network should be dominated by growing hubs.

For values of $p\in(0,1)$ the algorithm adds either (wih probability $p$) a single bond 
randomly, or it adds a collection of bonds (on average one bond) preferentially.  This 
grows simple networks of order $N$ and size $N-1$, typically not connected
unless acyclic.

In figure \ref{figure1} an example of a Barbasi-Albert network of order $122$ 
with $p=0$ is shown (left) and the right is a network of size $380$. The appearance of
hubs in these networks is clearly seen:  In the network on the left there are $5$
nodes of degrees exceeding $\sqrt{122}$, the largest of degree $31$,
and in the network on the right there are $3$ hubs of degrees exceeding $\sqrt{380}$,
the largest of degree $63$.

\subsection{Modified Barbasi-Albert networks}
Barabasi-Albert networks are relatively sparse networks.  A modification of
the algorithm can be introduced to grow denser networks.  For example,
one may replace step 3(b) by

\begin{enumerate}
\item[\bf ] 
\begin{enumerate}
\item[\bf 3(b).] With default probability $1-p$: Attach $x_n$ by adding bonds
$\edge{x_j}{x_n}$ with probability $q_j=\min\{\sfrac{\lambda\,k_j+A}{\sum_j k_j},1\}$
(where $\lambda$ and $A$ are non-negative parameters of the algorithm);
\end{enumerate}
\end{enumerate}

Since $k_j \ll \sum_j k_j$ in Barabasi-Albert networks, one may assume that
$\lambda k_j + A \leq \sum_j k_j$ for values of $\lambda$ and $A$ which 
are not too large (and so $q_j \leq 1$).  

\begin{figure}[h!]
 \centering
\includegraphics[height=9cm]{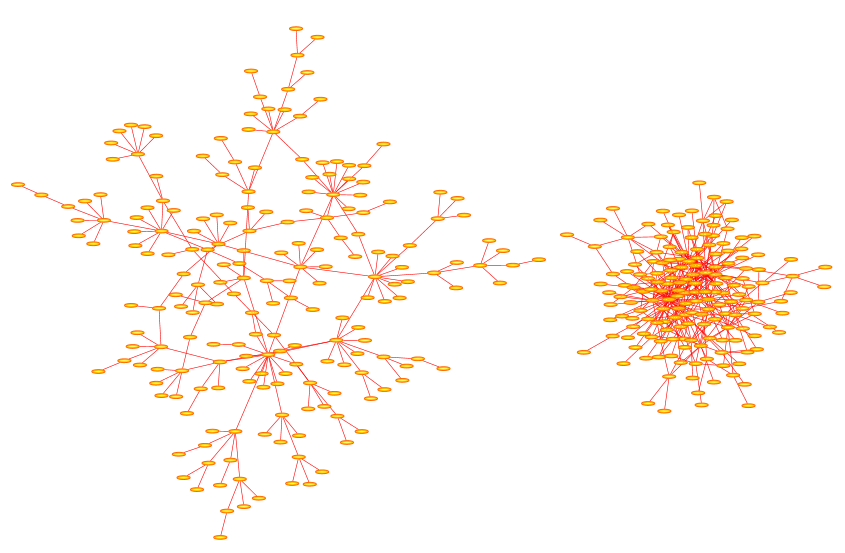}
\caption{{Modified Barabasi-Albert networks:}
The network on the left was grown with $\lambda=0.1$ to order $n=203$.  It has
two hubs of degrees $\{15,17 \}$ which exceed $\sqrt{203}$.
The network on the right was grown with $\lambda=1.5$ to order $n=172$.
This network contains hubs of degrees $\{15,15,16,17,19,27,33 \}$ exceeding
$\sqrt{172}$. In both cases the algorithm was implemented with $p=0$.  
The arrangement of nodes and bonds in
these networks was created using the prefuse force directed lay-out in Cytoscape 3.4.0
\cite{Cytoscape}.}
\label{figure2}
\end{figure}

In figure \ref{figure2} two examples of Modified Barabasi-Albert networks 
are shown, one a sparse network with $\lambda=0.5$,
$A=0$ and $p=0$, and the second a denser network with $\lambda=2.0$,
$A=0$ and $p=0$.  In both cases the algorithm was iterated $200$ times;
the sparse network has order $203$ and two hubs of degrees $\{15,17\}$, 
and the dense network has order $172$ with seven hubs of degrees
$\{15,15,16,17,19,27,33 \}$.

\subsection{Variant Barbasi-Albert networks}
A variant Barbasi-Albert algorithm can be introduced by changing
step 3(b) in the Barbasi-Albert algorithm to

\begin{enumerate}
\item[\bf ] 
\begin{enumerate}
\item[\bf 3(b).] With default probability $1-p$: Attach $x_n$ by adding bonds
$\edge{x_j}{x_n}$ with probability 
$q_j=\min\{\sfrac{k_j^\alpha+A}{\sum_j k_j},1\}$, (where
$\alpha$ and $A$ are non-negative parameters of the algorithm);
\end{enumerate}
\end{enumerate}

\begin{figure}[h!]
 \centering
\includegraphics[height=9cm]{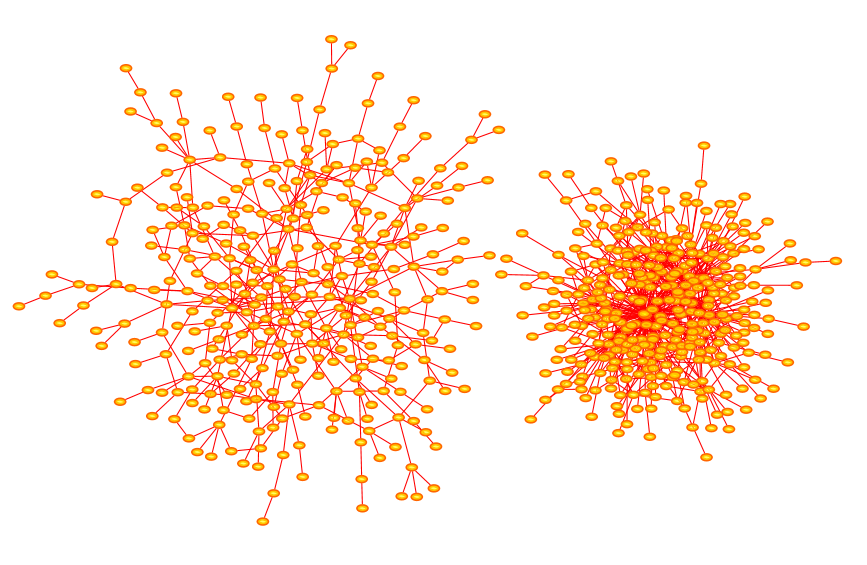}
\caption{{Variant Barabasi-Albert networks:}
The network on the left was grown using $\alpha=0.15$ and $A=0$ to
a total to $n=327$ nodes.    This graph is very sparse, and none of its nodes 
qualify as hubs.  The network on the right was grown to order $n=351$ with
$\alpha=1.15$ and $A=0$.  This is a dense network with several nodes
qualifying as hubs of degrees $\{22,24,26,42,43,116\}$.
 The arrangement of nodes and bonds in
these networks was created using the prefuse force directed lay-out in Cytoscape 3.4.0
\cite{Cytoscape}.}
\label{figure3}
\end{figure}

The effect of the parameter $\alpha$ is to increase the probability of adding bonds
to the hubs of the network if $\alpha>1$, and to decrease this probability if
$\alpha<1$.  In the case that $\alpha>1$ networks dominated by a single very
large hub are obtained (see figure \ref{figure3} (right network)), while networks with $\alpha<1$
are more sparse and not dominated by a few hubs (see figure \ref{figure3} (left network)).
The left network in figure \ref{figure3} was grown by putting $\alpha = 0.15$
and $A=0$ and has order $327$.  None of the nodes in this network has degree
which exceeds $\sqrt{327}$, and so none qualify as hubs.  
A denser network is obtained if $\alpha=1.15$ and $A=0$, as shown 
in figure \ref{figure3} on the right.  This network is dominated by
hubs of degrees $\{22,24,26,42,43,116\}$ and has order $351$. 

\subsection{Mean field theory for Modified Barabasi-Albert networks}
Let $k_j(n)$ be the degree of node $j$ after $n$ iterations.  A mean field calculation
of $k_j(n)$ is done by assuming that $k_j(n)$ is equal to its expected value for each $n$;
that is, $k_j(n) = \langle k_j (n) \rangle$ for each $j$ and $n$.

The modified Barabasi-Albert algorithm appends bonds to a network of order $n$ as
follows:  Step 3(a) is executed with probability $p$, and a bond (and the $(n+1)$-th node)
is appended with uniform probability on one of the $n$ existing nodes.  The probability
that node $j$ gets a bond in this way is $\sfrac{p}{n}$ and on average one bond is attached
with probability $p$.

If step 3(b) is done instead, then the expected number
of bonds added in the mean field is approximately $\sum_j \sfrac{\lambda\,k_j(n)+A}{\sum_j k_j(n)}
= \lambda + \sfrac{nA}{\sum_j k_j(n)}$. The total number of bonds in the network is
\begin{equation}
2E_n = \sum_j k_j(n)
\end{equation}
by handshaking.  Thus, the increment in the number of bonds when the next node is appended is
\begin{equation}
\Delta E_n = p + (1-p)\lambda + (1-p)\Sfrac{nA}{2E_n} .
\end{equation}
Approximate this by a differential equation
\begin{equation}
2E_n \sfrac{d}{dn} E_n = 2(p+(1-p)\lambda) E_n + (1-p)nA.
\end{equation}
This can be solved to obtain
\begin{equation}
E_n = \sfrac{n}{2}((p+(1-p)\lambda)+\sqrt{(p+(1-p)\lambda)^2+2(1-p)A}) = C\thin n,
\label{eqn7}  
\end{equation}
where $C$ is a function of $(p,\lambda,A)$ defined by this expression.  Notice that
$E_n$ grows linearly in $n$, so that Barabasi-Albert graphs will be necessarily sparse
as $n\to\infty$ (and by equation \Ref{eqn37} the scaling exponent is $\gamma>2$).

With each iteration the mean field value of $k_j(n)$ (the degree of the $j$-th node
after $n$ iterations) increments by
\begin{equation}
k_j(n+1) = k_j(n) + \Sfrac{p}{n} + \Sfrac{(1-p)(\lambda k_j(n) + A)}{2E_n} 
\end{equation}
since $2E_n = \sum_j k_j(n)=2C\thin n$, and since the probabilty of adding a bond to
node $j$ is $\sfrac{\lambda k_j(n)+A}{\sum_j k_j(n)}$.
This can again be approximated by a differential equation:  Take $n\to t$, a continuous
time variable, and let $k_j(n) \to k_j(t)$, the continuous mean field degree of node $j$.
Then
\begin{equation}
\sfrac{d}{dt} k_j(t) = \Sfrac{p}{t} + \Sfrac{(1-p)(\lambda k_j(t) + A)}{2Ct}.
\label{eqn6}  
\end{equation}
The initial condition is to assume that node $j$ is added at time $t_j$.
Putting $A=0$ and $\lambda=1$ gives $C=1$ and the equation
\begin{equation}
\sfrac{d}{dt} k_j(t) = \Sfrac{p}{t} + \Sfrac{(1-p)k_j(t)}{2t}
\end{equation}
which was also derived in reference \cite{BAJ99}.  In this event the solution is 
$k_j(t) = \sfrac{1+p}{1-p} (t/t_j)^{(1-p)/2}-\sfrac{2p}{1-p}$ (assuming the initial
condition $k_j(t_j)=1$).

More generally, equation \Ref{eqn6} can be cast in the general form
\begin{equation}
\sfrac{d}{dt} k_j(t) = \Sfrac{Q}{t} + \Sfrac{P}{t} k_j(t)
\end{equation}
where $Q= p + \sfrac{(1-p)A}{2C}$ and $P=\sfrac{(1-p)\lambda}{2C}$, with solution
\begin{equation}
k_j(t) = \left(1+\Sfrac{Q}{P}\right)\,(t/t_j)^P - \Sfrac{Q}{P}
\label{eqn12}   
\end{equation}
using again the intial condition $k_j(t_j)=1$.

The mean field degree distribution can be determined from this solution.  The probability
that node $j$ has degree $k_j(t)$ smaller than $\kappa$ at time $t$ is denoted by
$P[k_j(t) < \kappa]$.  Since $k_j(t) < \kappa$ if 
\[ \left(1+\Sfrac{Q}{P}\right)(t/t_j)^P < \kappa
\quad\hbox{or, equivalently,}\quad
t_j > t \left( \Sfrac{Q/P + \kappa}{1+Q/P} \right)^{-1/P} , \]
this is also the probability $P\left[(t_j/t) > \left( \sfrac{Q/P + \kappa}{1+Q/P} \right)^{-1/P}\right]$.
If the node $t_j$ is chosen uniformly from the $n$ available, then 
\begin{equation}
P[k_j(t) < \kappa] = P\left[(t_j/t) > \left( \sfrac{Q/P + \kappa}{1+Q/P} \right)^{-1/P}\right]
 = 1 -  \left( \Sfrac{Q/P + \kappa}{1+Q/P} \right)^{-1/P} .
\end{equation}
The mean field degree distribution is the derivative of this to $\kappa$:
\begin{equation}
P(\kappa) = P[k_j(t) = \kappa] = 
\sfrac{\partial}{\partial \kappa} P[k_n(t) < \kappa]
= \Sfrac{(P+Q)^{1/P}}{(P\kappa+Q)^{1+1/P}} .
\label{eqn11}  
\end{equation}

For large $\kappa$ this shows that the modified Barbasi-Albert network is scale-free with exponent
\begin{eqnarray}
\gamma &=& 1+ \sfrac{1}{P} = 1+ \Sfrac{2C}{(1-p)\lambda} \nonumber \\
&=& 1 + \Sfrac{((p+(1-p)\lambda)+\sqrt{(p+(1-p)\lambda)^2+2(1-p)A})}{(1-p)\lambda} .
\label{eqn15}   
\end{eqnarray}
Putting $A=0$ gives the exponent
\begin{equation}
\gamma = 3 + \Sfrac{2p}{(1-p)\lambda} .
\label{eqn16}   
\end{equation}
This is the mean field exponent of a modified Barabasi-Albert network.  For small $\lambda < 1$
the exponent is large, indicating a network  with few nodes (if any) of high degree.  
For large $\lambda>1$,  $\gamma \searrow 3^+$.  This is a lower bound 
on $\gamma$ for modified Barabasi-Albert networks.

If $\lambda = 1$, then the exponent $\gamma$ is given by
\begin{equation}
\gamma = 1 +\sfrac{1}{1-p} + \sfrac{\sqrt{1+2(1-p)A}}{1-p} .
\end{equation}
In this model one similarly finds that $\gamma \geq 3$, and in fact, if $p=0$, then
$\gamma = 2 +\sqrt{1+2A}$.  The parameter $A$ may be used to tune the
exponent $\gamma$ for any given $p$.

If both $\lambda=1$ and $A=0$, then the known expression for $\gamma$
for Barabasi-Albert networks is recovered, namely
\begin{equation}
\gamma = \Sfrac{3-p}{1-p}.
\end{equation}
Notice that $\gamma\geq 3$ and that $\gamma=3$ if $p=0$ \cite{BAJ99}.

The connectivity of modified Barabasi-Albert networks is given by
\begin{equation}
\langle k\rangle_n \simeq 
\SSfrac{\int_1^n k\, P(k) \, dk}{\int_1^n P(k) \, dk} 
\simeq \frac{2C}{2C-(1-p)\lambda} ,
\end{equation}
where $2C = ((p+(1-p)\lambda)+\sqrt{(p+(1-p)\lambda)^2+2(1-p)A})$.
Since $2-\gamma = 1-\sfrac{1}{P}$, equation \Ref{eqn3} gives
$\langle k\rangle_n \simeq \sfrac{1}{1-P}$.  Inserting the value of $P$ gives
the result above as well.

\begin{figure}[t!]
 \centering
\includegraphics[height=8cm]{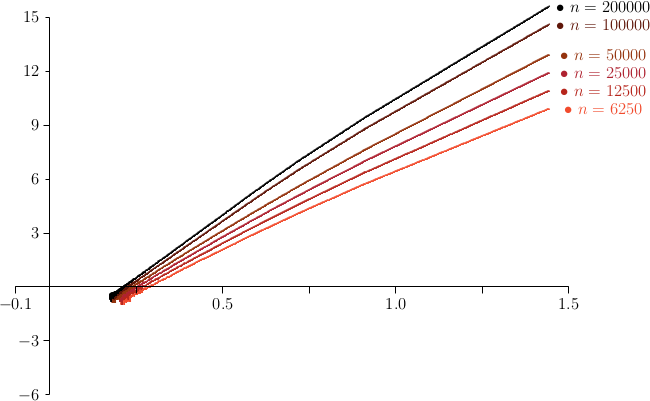}
\caption{{Scaling of Barabasi-Albert networks with $p=0$:}
Data on networks generated by the Barabasi-Albert algorithm with $p=0$.  In each case $100$
networks were grown and the average degree sequence $P_n(k)$ computed.  The curves above
are plots of $\log P_n(k)/\log (k+1)$ against $1/\log (k+1)$ for $n\in\{6250,12500,25000,\cdots,200000\}$.
Least squares fit to the data using a quadratic model gives the $y$-intercepts which 
averages to $3.026$.  This is very close to the value $\gamma=3$ predicted for the scaling
exponent in this model by the mean field approach.}
\label{figureBAgraph}
\end{figure}

In the figure \ref{figureBAgraph} the probability $P(k)$, that the degree
of a Barabasi-Albert network is equal to $k$, is examined by plotting
$\log P(k) / \log (k+1)$ against $1/\log(k+1)$ where $P(k)$ was estimated for
values $n\in\{6250,12500,25000,50000,100000,200000\}$ and for $p=0$.
The curves should intersect the vertical axis at $-\gamma$.   Least squares
fit of the data to quadratic curves gives $6$ estimates for $\gamma$, which
average to $\gamma = 3.026 \pm 0.076$, very close to the theoretical value $\gamma=3$
from equation \Ref{eqn16} (for $p=0$ and $\lambda=1$).

Data collected for the same values of $n$ and for $p=0.5$ cannot be successfully analysed
by regressions with quadratic curves, but cubic curves give the average value
$\gamma = 5.161 \pm 0.068$, which are not equal to but still fairly well approximated by $\gamma=5$
prediced by equation \Ref{eqn16} for $p=0.5$ and $\gamma=1$.

When $p=0.8$ the plots are strongly curved and extrapolation to estimate $\gamma$ is more
difficult.  In this case a different approach is needed.  Putting $\alpha = \sfrac{1}{2}$ in
equation \Ref{eqn2} gives
\begin{equation}
\log P(k) - \log P(\sfrac{1}{2}k) = - \gamma \log 2 + o(1)
\end{equation}
so that a plot of $\zeta(k) = (\log P(k) - \log P(\sfrac{1}{2}k))/\log 2 \to - \gamma$ as $k\to\infty$.
That is, plotting $\zeta(k)$ against $\sfrac{1}{k}$ gives a curve with $y$-intercept 
equal to $-\gamma$.  Better results are obtained when plotting against $\sfrac{1}{k}\log k$.
In this case a linear extrapolation gives $\gamma=11.67\pm0.41$ and a quadratic extrapolation
gives $\gamma=11.6\pm2.6$.  These results are close to the mean field prediction $\gamma=11$
for $p=0.8$.  Incidently, if $p=0.5$ then this kind of analysis show that $\gamma=5.47\pm0.14$
(linear extrapolation) or $\gamma=4.4\pm1.0$ (quadratic extrapolation), 
and if $p=0$, then the results are $\gamma=3.088\pm 0.022$ (linear extrapolation)
and $\gamma = 2.86\pm0.18$ (quadratic extrapolation).

If $\lambda=2$ and $p=A=0$ then the algorithm grows modified Barabasi-Albert networks
with $\gamma=3$ (the mean field estimate given by equation \Ref{eqn15}).  Estimating
$\gamma$ by plotting $\zeta(k)$ against $\sfrac{1}{k}\log k$ gives the estimate
$\gamma=3.019\pm 0.098$ (linear extrapolation) and $\gamma=2.62\pm 0.33$ (quadratic
extrapolation).

The connectivity of Modified Barabasi-Albert networks should converge quickly to a 
constant with increasing $n$ (by equation \Ref{eqn3}) since $\gamma>2$.  Computing
it for Barabasi-Albert networks (with $\lambda=1$ and $A=0$) gives
$\langle k\rangle_n \approx 3.16$ for $p=0$, $\langle k\rangle_n \approx 2.28$ for $p=0.5$
and $\langle k\rangle_n \approx 2.08$ for $p=0.8$, and for $n=12500$.  Increasing $n$
does not change these results.

\subsection{Mean field theory for Variant Barabasi-Albert networks}

In this model the increment in the number of bonds when the $(n+1)$-th node is appended
is given by
\begin{equation}
\Delta E_n = p + \Sfrac{(1-p)(\sum_j (k_j(n))^\alpha + A)}{2\, E_n} .
\end{equation}
Approximating this with a differential equation gives
\begin{equation}
2E_n\sfrac{d}{dn} E_n = 2p E_n + (1-p)nA + (1-p)\sum_j (k_j(n))^\alpha .
\end{equation}
The right hand side can be approximated as follows:  For $\alpha>1$ the algorithm
should grow dense networks with nodes of high degree.  Assuming that $k_j(n) \approx
k_\ell(n)$ for all $\ell$ shows that $\sum_j (k_j(n))^\alpha \approx n (k_j(n))^\alpha
\approx n\left(\sfrac{1}{n} \sum_j k_j (n) \right)^\alpha
= n^{1-\alpha} (2E_n)^\alpha$.  Using this approximation gives
\begin{equation}
2E_n\sfrac{d}{dn} E_n \approx  2p E_n + (1-p)nA + (1-p)n^{1-\alpha}(2E_n)^\alpha.
\label{eqn21}  
\end{equation}
If $A=p=0$, then the differential equation can be solved directly to obtain
$E_n \simeq 2^{(\alpha-1)/(2-\alpha)} n$, provided that $\alpha>1$. This shows that
$E_n$ is linear in $n$, which may be expected if $\alpha$ is not too much larger 
than $1$.

Numerical experimentation shows that $E_n$ grows linearly in $n$ for values
of $\alpha$ not too much larger than $1$.  For example, if $p=0.5$, $A=1$ and $\alpha=1$ then 
$\sfrac{1}{n} E_n \to 1.207\ldots$, if $\alpha=1.5$ then
$\sfrac{1}{n} E_n \to 1.539\ldots$, but if $\alpha=2$ then $\sfrac{1}{n}E_n$ increases
slowly with $n$.  Similarly, if $p=0$, and $A=1$, then, if $\alpha=1$,
$\sfrac{1}{n} E_n \to 1.366\ldots$, and if $\alpha=1.5$,
$\sfrac{1}{n} E_n \to 2.399\ldots$, but if $\alpha=2$ then $\sfrac{1}{n}E_n$ increases
slowly with $n$ and for even larger values of $n$ this growth accelerates. 

The recurrence for the degree of the $j$-th node
may be approximated by a differential equation similar to equation \Ref{eqn6}:
Assuming that $E_n = D n^\beta$, replacing $n\to t$ (a continuous time variable), gives
the recurrence
\begin{equation}
k_j (t+1) = k_j(t) + \Sfrac{p}{t} +  \Sfrac{(1-p)((k_j(t))^\alpha + A)}{2D t^{\beta}}.
\end{equation}
This can be approximated by the differential equation
\begin{equation}
\sfrac{d}{dt} k_j(t) = \Sfrac{p}{t} + \Sfrac{(1-p)((k_j(t))^\alpha + A)}{2D t^{\beta}}.
\label{eqn23}  
\end{equation}
 If $\alpha=1$ and $\beta=1$ then the solution of this equation gives
the Barabasi-Albert case with $\gamma=3$.  Proceed by considering the case $A=p=0$
and the initial condition $k_j(t_j)=1$.  Assume that $\alpha = 1 + \epsilon$.  Then the 
equation becomes
\begin{equation}
\Sfrac{2Dt^\beta}{k_j(t)}\, \sfrac{d}{dt} k_j(t) = (k_j(t))^\epsilon .
\end{equation} 
A perturbative approach for small $\epsilon$ can be done by expanding
$(k_j(t))^\epsilon = \hbox{exp} (\epsilon \log k_j(t)) = 1 + \epsilon \log k_j(t) + \sfrac{1}{2}\epsilon^2
\log^2 k_j(t) + \cdots$.  Truncating this at $O(\epsilon^2)$ and putting $g(t) = \log k_j(t)$
gives the differential equation
\begin{equation}
2Dt^\beta \sfrac{d}{dt} g(t) = 1 + \epsilon g(t) + \sfrac{1}{2} \epsilon^2 g^2(t) .
\end{equation}
Using the initial condition $g(t_j) = \log k_j(t_j) = 0$ the solution of this equation is
\begin{equation}
\epsilon\, g(t) = 
\cases{
-1 + \tan \left( \Sfrac{\pi}{4}  + \Sfrac{\epsilon}{4D} \log (\Sfrac{t}{t_j} ) \right),&\quad\hbox{if $\beta=1$}; \\
-1 + \tan \left(  \Sfrac{\pi}{4} 
+ \Sfrac{\epsilon}{4D(\beta-1)}(t_j^{1-\beta}-t^{1-\beta} ) \right),&\quad\hbox{if $\beta>1$}.
}
\end{equation}
In the case $\beta>1$ suppose that $\delta=\beta -1$ and that $\delta$ is small.
Then approximate
\[ \hspace{-2.5cm}
t_j^{1-\beta}-t^{1-\beta}
= e^{-\delta \log t_j} - e^{-\delta \log t} \approx \delta \log \left(\sfrac{t}{t_j}\right) 
- \sfrac{1}{2}\delta^2  \log \left(\sfrac{t}{t_j}\right)\log \left(t t_j\right) + O(\delta^3). \]
With this approximation the solution for $g(t)$ above can be expanded in $\epsilon$
and $\delta$ to give the first order approximations
\begin{equation*}
\hspace{-2.5cm}
g(t) \simeq 
\cases{
\Sfrac{1}{2D} \log ( \Sfrac{t}{t_j} ) + \Sfrac{\epsilon}{8D^2}\log^2 \Sfrac{t}{t_j},&\quad\hbox{if $\beta=1$}; \\
\Sfrac{1}{2D} \log ( \Sfrac{t}{t_j} ) + \Sfrac{\epsilon}{8D^2}\log^2 \Sfrac{t}{t_j}
- \Sfrac{\delta}{4D^2}\left(D \log^2 ( \Sfrac{t}{t_j} )+\log t_j \, \log ( \Sfrac{t}{t_j} )   \right),&\quad\hbox{if $\beta>1$}.
}
\end{equation*}

Proceed by solving the above quadratics for $\log(\sfrac{t}{t_j})$ in terms of $g(t)$.  Expand the
solution in $\epsilon$ and $\delta$ and keep only the first few terms.  In the case that $\beta=1$
this gives
\begin{equation}
\log ( \Sfrac{t}{t_j} ) \approx 2D\,g(t) - \epsilon D \, g^2(t) .
\end{equation}
Since $g(t) = \log k_j(t)$, the probability that $k_j(t) < \kappa$ is given by
\begin{equation}
P[k_j(t) < \kappa] = P\left[\Sfrac{t_j}{t} > \kappa^{\epsilon D\log \kappa - 2D} \right]
\approx 1 - \kappa^{\epsilon D\log \kappa - 2D} .
\end{equation}
Taking the derivative to $\kappa$ gives the distribution function in the case that $\beta=1$:
\begin{equation}
P(k) \sim D(2-D\epsilon\log k)\, k^{-1-2D+D\epsilon\,\log k} .
\end{equation}
These networks are thus not scale-free.  For small values of $k$ the $\log k$ terms are slowly
varying, and the networks will appear to be scale-free with $\gamma=1+2D$.  However,
with increasing $k$ the exponent reduces in value and the connectivity of the network will
become dependent on $k$ in the way seen in equation \Ref{eqn3} for small values of $\gamma$.

Notice that if $D=1$ and $\epsilon=0$ (or $\alpha=1$), then the above reduces to $P(k) \sim k^{-3}$,
as expected for Barabasi-Albert networks.

If $\beta>1$, then a similar approach to the above may be considered.  Solving the expression
for $g(t)$ above for $\log( \sfrac{t}{t_j} )$ and keeping only terms to $O(\epsilon)$ and $O(\delta)$ gives
\begin{equation}
\log ( \Sfrac{t}{t_j} ) \approx 2D\,g(t) - \epsilon D \, g^2(t)  + \delta(2D^2 g^2(t) + g(t) \log t_j) .
\end{equation}
This shows that
\begin{eqnarray*}
P(k_j(t) < \kappa) &= P\left(\sfrac{t_j}{t} > \kappa^{\epsilon D\log \kappa - 2D
- 2D^2\delta\log \kappa - \delta\log t_j}  \right) \\
&\approx 1 -\kappa^{\epsilon D\log \kappa - 2D
- 2D^2\delta\log \kappa - \delta\log t_j} .
\end{eqnarray*}
This shows that
\begin{equation}
P(k) \sim (2D(1+2D\delta\log k - \epsilon \log k))\,
k^{-1-2D-\delta\log t_j - D(2D\delta-\epsilon)\log k} .
\end{equation}
This gives an effective exponent $\gamma_k = 1+2D+\delta\log t_j + D(2D\delta-\epsilon)\log k$
which decreases in size if $2D\delta - \epsilon < 0$ and increases in size if $2D\delta-\epsilon>0$.
Since $\delta=\beta-1$ and $\epsilon=\alpha-1$, and for small $\alpha$ numerical simulations
show that $\beta\approx 1$, it is normally the case that $2D\delta - \epsilon <0$.  This means
that the networks will first appear scale-free with constant connectivity until $k$ becomes
large enough in which case the connectivity will increase with $k$, as seen above.

\begin{figure}[t!]
 \centering
\includegraphics[height=8cm]{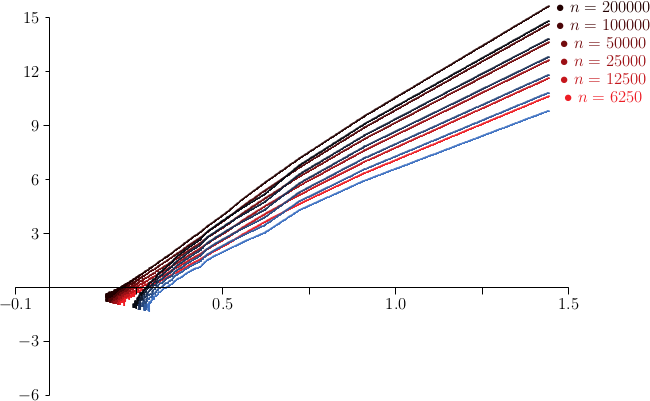}
\caption{{Variant Barabasi-Albert networks with $p=0$:}
Data on networks generated by the Barabasi-Albert algorithm with $p=0$ and $\alpha=1.1$
(red curves) and $\alpha=0.5$ (blue curves).  In each case $100$ networks were grown and 
the average degree sequence $P_n(k)$ computed.  The curves above
are plots of $\log P_n(k)/\log (k+1)$ against $1/\log (k+1)$ for $n\in\{6250,12500,25000,\cdots,200000\}$.
}
\label{figureVBAgraph}
\end{figure}

\subsection{Numerical results on Variant Barabasi-Albert networks}

In figure \ref{figureVBAgraph} data for networks with $p=0$ and $\alpha=1.1$ and $\alpha=0.5$
is shown.  Since $\alpha=1.1$ is still very close to $1$, the results above show that these
networks should still appear scale-free, and with connectivity a constant.  This is indeed the
case.  For $n=6250$ the data gives $\langle k \rangle_n = 3.149$, and increasing
$n$ to $n=200000$ gives $\langle k \rangle_n = 3.176$.  That is, the connectivity of the
networks are insensitive to $n$ over this range.  Least squares fits to the curves with quadratic
polynomials in order to determinate the value of $\gamma$ give the average
$\gamma = 2.857\pm 0.068$.  This result is consistent with a constant value of 
the connectivity of networks of these size ranges.  With increasing $n$, it is expected
that $\gamma $ will decrease in value (that is, the value given here is an effective
value), and eventually, the connectivity will start to increase.

Networks generated with $p=0$ and $\alpha=0.5$ turned out to be sparse with
low connectivity.  For example, for $n=100000$, the connectivity is $\langle k \rangle_n=1.036$
and this decreases even further for $n=200000$, where $\langle k \rangle_n = 1.020$.
Attemps to extract an exponent $\gamma$ from the data for these networks were not succesful,
the regressions did not settle on a value, but are strongly dependent on $n$.  Notice that
the mean field analysis above does not apply to networks with $\alpha < 1$.

Putting $\alpha=2$ gives networks with average connectivity which increases with
$n$.  For example,  if $n=100$, then $\langle k \rangle_n = 43$, for $n=500$, 
$\langle k \rangle_n = 260$ and for $n=1000$, $\langle k \rangle_n = 527$.  
On the other hand, if $\alpha = \sfrac{3}{2}$, then  $\langle k \rangle_n = 3.08$ if $n=100$,
$\langle k \rangle_n = 3.27$ if $n=500$, and $\langle k \rangle_n = 3.31$ if $n=1000$, 
and it appears that for small values of $n$ the connectivity does not change quickly
with increasing $n$.

\section{Duplication-Divergence networks}

Biological models of protein evolution are usually presented in terms of two
processes, namely (1) a \textit{duplication event} involving a gene sequence in DNA, 
and (2) a \textit{(random) mutation} of duplicated genes which then drift
from one another in genetic space \cite{W03,BLW04,HZ05}.   The mutations of
duplicated and mutated genes change the proteome and the network of protein
interactions:  If the protein is self-interacting, then the duplicated proteins interact,
and the mutated genes code for proteins with altered interactions (some gained,
others weakened or lost) with other proteins. 

The Duplication-Divergence algorithm models these processes in order to grow a
network, and was used in order to estimate the rates of duplication and mutation
in the protein interaction networks \cite{VF02}.  There is a rich and large 
literature reporting on modeling protein interaction networks using models
which include processes of duplication and divergence \cite{EI08,KBL04,TD11,NSRA12}.

Since proteomic networks appear to be scale-free \cite{E06,KW06}, it seems likely that
duplication and divergence processes should grow scale-free networks and that this
should also be seen in computer algorithms which grow networks using
duplication and divergence elementary moves.  Duplication can be implemented by
selecting nodes and duplicating them, and their incident bonds, in a network.
Divergence is implemented by altering the bonds incident on particular nodes,
namely either by deleting, adding or moving bonds.  In the Duplication-Divergence
algorithm these moves are implemented by selecting  nodes uniformly for duplication
to progenitor-progeny pairs, 
and by deleting bonds incident to either the progenitor node or its progeny. 
Notice that since nodes of high degree have a larger probability of being adjacent
to a node selected for duplication, these nodes have a larger probability of 
receiving new bonds in the duplication process -- in this way there are events
of preferential attachment in this algorithm \cite{EL03,PS03}.

\begin{figure}[t!]
 \centering
\includegraphics[height=2.5in]{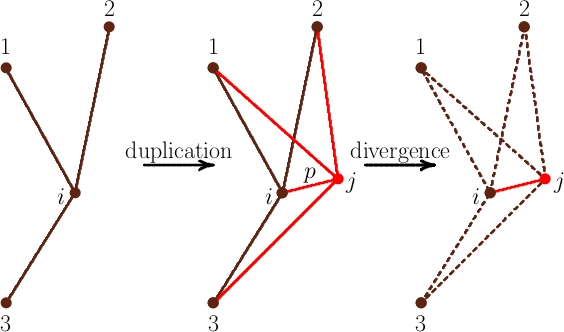}
  \caption{{The Duplication-Divergence algorithm:}
Duplication-Divergence iterations:  A node $i$ and its incident bonds are duplicated to 
create a node $j$ with its incident bonds.  The bond
$\langle{i}{\sim}{j}\rangle$ is added with probability $p$.  In the divergence
step one of the pair of bonds $(\edge{i}{m},\edge{j}{m})$ is deleted with
probability $q$, for each value of $m\in\{1,2,3\}$.}
\label{figure5DD}
\end{figure}

The basic elementary move of the Duplication-Divergence algorithm is illustrated
in figure \ref{figure5DD}.  The algorithm is implemented as follows.

\vspace{1mm}
\noindent{\bf Duplication-Divergence algorithm:}
\begin{enumerate}
\item[\bf 1.] Initiate the network with one node $x_0$ and apply the following steps iteratively;
\item[\bf 2.] Duplication: Choose a node $\upsilon$ uniformly and duplicate by creating
node $\upsilon^\prime$;
\item[\bf 3.]  For all bonds $\edge{w}{\upsilon}$ incident with
$\upsilon$, add the bonds $\edge{w}{\upsilon^\prime}$;
\item[\bf 4.] With probability $p$ add the bond $\edge{\upsilon}{\upsilon^\prime}$;
\item[\bf 5.] Divergence: delete one bond of the pair $\{ \edge{w}{\upsilon}, \edge{w}{\upsilon^\prime}\}$
 incident with $\upsilon$ or with its duplicated node $\upsilon^\prime$ with probability $q$
(for each $w$ adjacent to both $\upsilon$ and $\upsilon^\prime$ independently);
\item[\bf 6.] Stop the algorithm when a network of order $N$ is grown.
\qed
\end{enumerate}
\vspace{5mm}

The algorithm has two parameters $(p,q)$.  

The parameter $p$ is the probability that the protein corresponding to the
progenitor node $\upsilon$ is self-interacting.  If it is (with probability $p$)
then the bond $\edge{\upsilon}{\upsilon^\prime}$ is added to the network
and it represents the interaction between $\upsilon$ and $\upsilon^\prime$.

The parameter $q$ controls the model of divergence in this algorithm.  As
$\upsilon$ and $\upsilon^\prime$ diverge from one another,  one bond in 
each pair of bonds incident with $\upsilon$ and $\upsilon^\prime$  is lost
independently, with probability $q$.  The result is that the network mutates
as bonds (interactions) are lost (while they are created by the duplication
process).

A slightly modified algorithm is found by changing step 5 in the algorithm to
find a modification of the Duplication-Divergence algorithm which assumes that
one of the duplicated pair mutates, while the other remains stable.

\begin{enumerate}
\item[\bf ] 
\begin{enumerate}
\item[\bf 5.] Divergence: Consider all bonds $\edge{w}{\upsilon^\prime}$ incident
with the duplicated node $\upsilon^\prime$ and delete these independently
with probability $q$.
\end{enumerate}
\end{enumerate}

The Duplication-Divergence algorithm tends to grow disconnected networks, while the
Modifed Duplication-Divergence algorithm is more likely to grow networks with a
single component (that it, connected networks).

\subsection{Mean field theory for Duplication-Divergence networks}
Let $k_j(n)$ be the degree of node $j$ after $n$ iterations.  The algorithm appends
nodes by duplicating them (the probability that a node $\upsilon$ is duplicated
in a network of order $n$ is $\frac{1}{n}$), adds bonds by inserting a bond
between a node and its duplicate with probability $p$, and remove bonds 
by selecting one bond between node-duplicate pairs and other nodes independently
and deleting it with probability $q$.  Let $2\thin E_n = \sum_j k(n)$ be twice the total 
number of bonds after $n$ iterations. Then, if $ k_j(n)$ is the
degree of node $j$ at time $n$, and node $j$ is duplicated, the number of
bonds in the network $E_n$ increases in the mean field by
\begin{equation}
E_{n+1} = E_n + p + k_j(n) - q\, k_j(n) .
\label{eqn28}  
\end{equation}
This follows since $k_j(n)$ bonds are created in the duplication
move in the mean field, and another bond is created between the $j$-th node 
and its duplicate with probability $p$.  The number of deleted bonds in the mean 
field is $q\, k_j(n) $.

Notice that $2\thin E_n = \sum_j k_j(n) = n\thin a_n$ where $a_n = \langle k_j(n) \rangle$ is
the average degree.  In the mean field approximation one substitutes $ k_j(n)$ in 
the recurrence \Ref{eqn28} by its network average $a_n$.  Then equation \Ref{eqn28}
can be casted as a recurrance for $a_n$:
\begin{equation}
(n+1)\,a_{n+1} = n\, a_n + 2p + 2(1-q)\, a_n .
\end{equation}
Let $n\to t$, where $t$ is a continuous time variable, and approximate this 
recurrence by the differential equation
\begin{equation}
t \sfrac{d}{dt} a_t = 2p + (1-2q)\, a_t .
\label{eqn34}   
\end{equation}
The initial condition is $a_1=1$, and this has solution
\begin{equation}
a_t = \Sfrac{1-2(q-p)}{1-2q} \, t^{1-2q} -\Sfrac{2p}{1-2q} .
\label{eqn35}   
\end{equation}
Since $E_n \simeq \sfrac{1}{2} n\, a_n$, it follows that
\begin{equation}
E_n 
= \Sfrac{1-2(q-p)}{2(1-2q)}\, n^{2(1-q)} - \Sfrac{pn}{1-2q} .
\label{eqn36}   
\end{equation}
Comparison to equation \Ref{eqn37} shows that, if $q<\sfrac{1}{2}$, 
\begin{equation}
\gamma = 1+2q .
\end{equation}
In this case $E_n = O(n^{2(1-q)}) + O(n)$ and that while $2(1-q)>1$, the term $O(n)$ is a 
strong correction to the growth in $E_n$ for even large values of $n$.  In
other words, the degree distribution $P(k)$ of the network will be strongly
corrected from the powerlaw distribution in equation \Ref{eqn1}.

If $q=\sfrac{1}{2}$, then by solving equation \Ref{eqn34}, $a_t = 1 + 2p\log t$ 
(so that $a_1 = 1$).  Since $E_n = \sfrac{1}{2} n \, a_n$, this shows that
\begin{equation}
E_n = \sfrac{1}{2}n + pn \log n, \quad\hbox{if $q=\sfrac{1}{2}$.}
\end{equation}
In this case $\gamma=2$ by equation \Ref{eqn37}, but notice the
subtle domination of the $n\log n$ term.  In numerical work this
will be very hard to see.

The case $q>\sfrac{1}{2}$ is considered by noting that $a_t \simeq \sfrac{2p}{2q-1}$ as
$t\to\infty$.  This shows that
\begin{equation}
E_n \simeq \Sfrac{p\,n}{2q-1}, \quad\hbox{if $q>\sfrac{1}{2}$.}
\label{eqn43}   
\end{equation}
This shows that $\gamma\geq 2$ by equation \Ref{eqn37}.  

Putting the above together gives
\begin{equation}
\gamma \; 
\cases{
=1+2q, &\hbox{if $q \leq \sfrac{1}{2}$}; \\
\geq 2,  &\hbox{if $q > \sfrac{1}{2}$}.
}
\label{eqn43x}   
\end{equation}
with a logarithmic correction if $q=\sfrac{1}{2}$.

Comparing the coefficient in equation \Ref{eqn37} with equation \Ref{eqn43}
gives a refined estimate $\gamma =1 + \sfrac{2p}{1+2p-2q} \geq 2$, provided
that $2q<1+2p$.  For example, if $q=0.75$ then $p>0.25$.   However, numerical
work shows this estimate to be too small, and estimating $\gamma$
in this regime for this model remains an open question.

\begin{figure}[t!]
 \centering
\includegraphics[height=9cm]{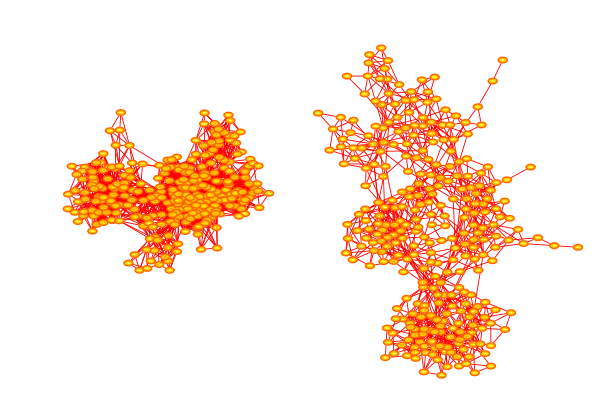}
\caption{{Duplication-Divergence network:}
The network on the left is a network generated with $p=1$ and
$q=0.40$.  It has order $300$ and it has $114$ nodes with degrees
exceeding $\sqrt{300}$ and so qualify as hubs.  The largest few of these
hubs have degrees $\{ 43,45,47,47,50\}$.  The network
on the right is similarly a network generated with $p=1$ and
$q=0.60$.  It is more extended but has only one node of degree equal to one.
Its order is $300$, and it has $5$ nodes of degrees $\{18,18,19,20,23\}$ which
qualify as hubs.  Networks generated with the Modified Duplication-Divergence
algorithm have a similar appearance, with the exception that more nodes
of degree $1$ are seen.  The arrangement of nodes and bonds in
these networks was created using the prefuse force directed lay-out in Cytoscape 3.4.0
\cite{Cytoscape}.}
\label{figure4}
\end{figure}

The power law decrease in $P(k)$ in  equation \Ref{eqn1} is only asymptotic
for this algorithm; and there should be corrections in particular for $q<\sfrac{1}{2}$.  
From the results above the average connectivity can be computed:  
Since $E_n = \sfrac{1}{2}n\, \langle k_j(n) \rangle$, 
\begin{equation}
\langle k \rangle_n \simeq
\cases{
\Sfrac{1-2(q-p)}{1-2q}\, n^{1-2q} - \Sfrac{2p}{1-2q}, &\hbox{if $q<\sfrac{1}{2}$}; \\
2p\log n + 1, &\hbox{if $q=\sfrac{1}{2}$}; \\
\hbox{Constant}, &\hbox{if $q>\sfrac{1}{2}$}.
}
\label{eqn44}   
\end{equation}
From these results $P(k)$ can be calculated.  Since $\langle k \rangle_n
\simeq \int_1^n k\, P(k)\, dk$, it follows that $\sfrac{d}{dn} \langle k \rangle_n
= n\, P(n)$.  Thus, using this approach gives
\begin{equation}
P(k) \sim
\cases{
(1-2(q-p)) \, k^{-1-2q}, &\hbox{if $q<\sfrac{1}{2}$}; \\
2p\, k^{-2} , &\hbox{if $q=\sfrac{1}{2}$}; \\
C_0\, k^{-\gamma } , &\hbox{if $q>\sfrac{1}{2}$},
}
\end{equation}
where the case $q>\sfrac{1}{2}$ is unknown since the dependence of the
exponent $\gamma$ on the parameters $(p,q)$ is not known.  Notice the
change in behaviour at the critical value $q=\sfrac{1}{2}$; this was already 
observed numerically in reference \cite{VF02}.

The modified Duplication-Divergence algorithm has the same recurrence 
\Ref{eqn35}, and so the values for $\gamma$ and relations for $\langle k \rangle_n$
and $P(k)$ remain unchanged for this algorithm.  Notice that this implementation
preserves the degree of the selected node, and tends to give a duplicated node
with lower degree (while the (unmodified) implementation tends to lower
the degrees of both the selected and duplicated nodes).  As a result, networks
generated with the modified algorithm have, on average, more nodes of
degree equal to one (and so appear more tree-like).

\begin{figure}[t!]
 \centering
\includegraphics[height=8cm]{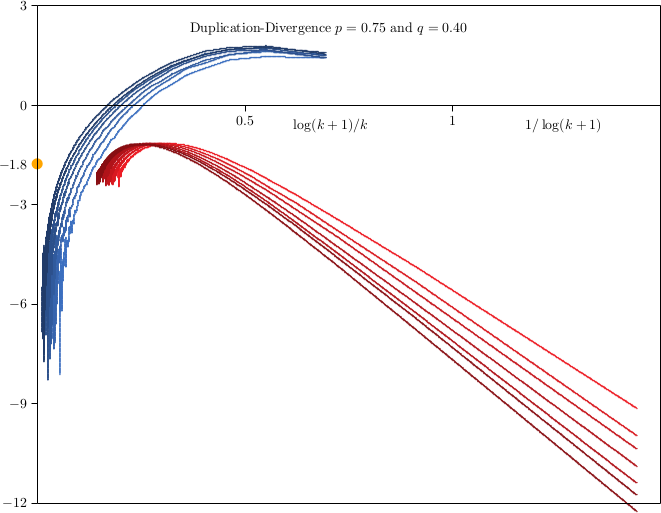}
\caption{{The distribution of degrees in Duplication-Divergence networks with
$p=0.75$ and $q=0.40$:}
Data on networks generated by the Duplication-Divergence algorithm.  In each case $100$ networks 
were grown and the average degree sequence $P_n(k)$ computed.  The curves on the right 
are plots of $\log P_n(k)/\log (k+1)$ against $1/\log (k+1)$ for $n\in\{3125,6250,12500,\cdots,200000\}$,
while those on the left are plots of $(\log P(2k) - \log P(k))/\log 2$ as a function of 
$\log(k+1)/k$. The mean field estimate for the exponent $\gamma$ is marked at
$-\gamma=-1.8$ on the left hand axis.  The strong correction to scaling evident in these curves
makes it difficult to extrapolate to the mean field value for $\gamma$.
}
\label{figureVaz40}
\end{figure}

\subsection{Numerical results on Duplication-Divergence networks}

In figure \ref{figure4} two networks grown with the Duplication-Divergence
algorithm are shown.  Both networks were grown with $p=1$ and have 
order $300$. The network on the left was grown with divergence parameter
$q=0.4$, and that on the right, with the higher mutation rate $q=0.6$.   

In figure \ref{figureVaz40} data for networks grown with $p=0.75$
and $q=0.4$ are shown.  The curves on the right were obtained by plotting
$(\log P(k))/\log(k+1)$ averaged over $100$ networks of sizes
$\{3125,6250,12500,25000,50000,100000,200000\}$ against $1/\log(k+1)$.
The mean field value of $\gamma$ is denoted by the bullet on the left-hand
axis.  These data show that convergence to this value is very slow -- this indicates
strong corrections to scaling arising in equation \Ref{eqn36}.

An alternative approach is to estimate $\gamma$ by plotting $\zeta(k) = 
(\log P(2k) - \log P(k))/\log 2$ as a function of $\log(k+1)/k$ (see equation
\Ref{eqn2} with $\alpha=2$).  The results are also strongly curved data
(left in figure \ref{figureVaz40}), and while the results are not inconsistent 
with the mean field value $\gamma \approx 1.9$ in this model, 
however, it  seems difficult to extrapolate these curves to a limiting 
value of $\gamma$.

\begin{figure}[t!]
 \centering
\includegraphics[height=8cm]{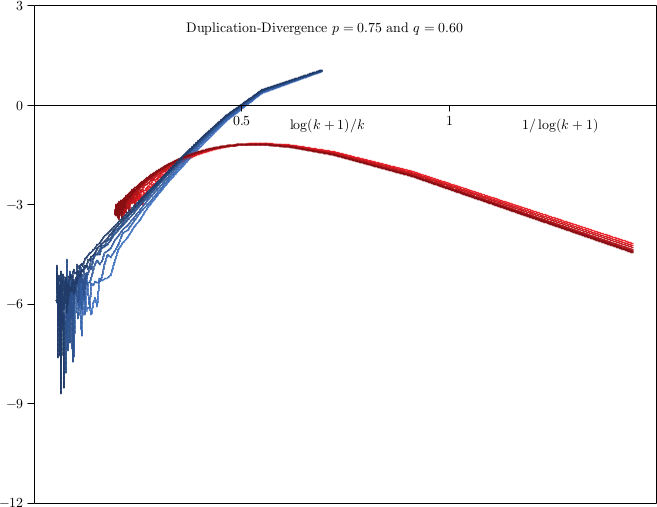}
\caption{{The distribution of degrees in Duplication-Divergence networks with
$p=0.75$ and $q=0.60$:}
Data on networks generated by the Duplication-Divergence algorithm.  In each case $100$ networks 
were grown and the average degree sequence $P_n(k)$ computed.  The curves on the right 
are plots of $\log P_n(k)/\log (k+1)$ against $1/\log (k+1)$ for $n\in\{3125,6250,12500,\cdots,200000\}$,
while those on the left are plots of $(\log P(2k) - \log P(k))/\log 2$ as a function of 
$\log(k+1)/k$.  Each of these curves can be extrapolated by a quadratic least squares fit to
obtained estimates of $\gamma$.  This gives the estimates $\gamma_n$ for $n=3125\times 2^\ell$
for $\ell=0,1,2,\ldots, 6$.  Extrapolating the $\gamma_n$ to $n=\infty$ by a least squares
fit $\gamma_n = \gamma + A/n$ gives $\gamma\approx 7.4$.
}
\label{figureVaz60}
\end{figure}

If $q=0.60>\sfrac{1}{2}$ then the results in figure \ref{figureVaz60} are seen.
The curves of $\zeta(k) = (\log P(2k) - \log P(k))/\log 2$ as a function of $\log(k+1)/k$ 
have straightened considerably, and each can be extrapolated by a quadratic
least squares to obtain an estimate $\gamma_n$ for each value of
$n=3125\times 2^\ell$ (for $\ell=0,1,2,\ldots,6$).  This gives estimates
$\{9.68,8.52,7.99,7.95,7.82,7.58,7.05\}$ which can be extrapolated by a
least squares fit of $\gamma_n = \gamma + A/\log n$, giving the estimate
$\gamma \approx 2.87$, which is slightly larger than the value predicted
by the mean field formula $\gamma = 1+ \sfrac{2p}{1+2p-2q}$
(see the paragraph following equation \Ref{eqn43x}).  This suggests that 
the approach to limiting behaviour in this model is quite slow, consistent with
the remarks after equation \Ref{eqn43x} in the previous section.

The average connectivity $\langle k \rangle_n$ is expected to behave according
to equation \Ref{eqn44}.  In table \ref{table1} $\langle k \rangle_n$ is listed
for $p=0.75$ and $q=0.40$, $q=0.50$ and $q=0.60$.  If $q=0.4$, then equation
\Ref{eqn44} suggests that $\langle k \rangle_n \simeq 8.5\, n^{0.2}$.  Computing
$\langle k \rangle n \times n^{-0.2}$ from the data in table \ref{table1} gives
$\{5.18,5.45,5.65,5.91,5.96,6.01,6.12\}$.  Plotting these results against $1/\log n$
and then linearly exptrapolating as $n\to\infty$ gives $7.98$, close to the 
value of $8.5$ predicted in equation \Ref{eqn44}.

If $q=0.5$, then equation \Ref{eqn44} suggests that $\langle k \rangle_n
\simeq 1.5 \log n$ since $p=0.75$.  Dividing the results in table \ref{table1}
by $\log n$ for each value of $n$ gives the results $\{1.42,1.44,1.44,1.42,1.43,1.46,1,45 \}$.
The average of this is close to the predicted value of $1.5$.

Finally, if $q=0.6$ then the data appear to approach a constant. Extrapolating 
these results using the model $A + B/\log(n)$ gives the estimated limiting
value $8.72$.  By equation \Ref{eqn3} this indicates that $\gamma=2.13$,
a value which is quite close to $2.15$, the value predicted by 
the formula $\gamma = 1+ \sfrac{2p}{1+2p-2q}$ in the paragraph following equation
\Ref{eqn43x}.

\begin{table}[t!]
\centering
\caption{Connectivity data for Duplication-Divergence Networks.}
      \begin{tabular}{lccc}
        \hline
       $n$    &  $q=0.4$  & $q=0.5$  & $q=0.6$ \\ \hline
        $3125$ & $25.9$ & $11.4$ & $5.93$ \\
        $6250$ & $31.3$ & $12.6$ & $6.14$ \\
      $12500$ & $37.3$ & $13.6$ & $6.33$ \\
      $25000$ & $44.8$ & $14.4$ & $6.55$ \\
      $50000$ & $51.9$ & $15.5$ & $6.64$ \\
    $100000$ & $60.1$ & $16.8$ & $6.75$ \\
    $200000$ & $70.3$ & $17.7$ & $6.88$ \\ \hline
      \end{tabular}
\label{table1}
\end{table}

\section{Sol\'e evolutionary networks}

The Sol\'e model \cite{SPSK02,PS03}
modifies  Duplication-Divergence model by using duplication and network 
rewiring as the basic elementary moves.  As before, the duplication of nodes is an implementation
of gene duplication, and the network rewiring is based on the loss and gain of protein 
interactions in the bulk of the network \cite{BAJ00}.   Thus, the algorithm grows
networks based on a model of gene duplication and the rewiring of 
protein interactions; both these processes drive the evolution of the interactome.

The elementary move of the algorithm is as follows:  A node in the network is chosen 
uniformly and randomly, and duplicated to form a progenitor-progeny pair.  The
progeny will have the same interactions as the progenitor.  This network is updated in the
rewiring step which has two parts:  Bonds incident with the progeny protein are deleted
with probability $\delta$, and new bonds are added in the network between nodes (excluding
the progenitor protein) are created with probability $\alpha$.  This implementation differs in
two ways from the Duplication-Divergence algorithm.  In the Sol\'e model there 
are no self-interacting nodes, and the formation of new bonds in the rewiring steps
only occurs in the Sol\'e model.

\begin{figure}[t!]
 \centering
\includegraphics[height=1.9in]{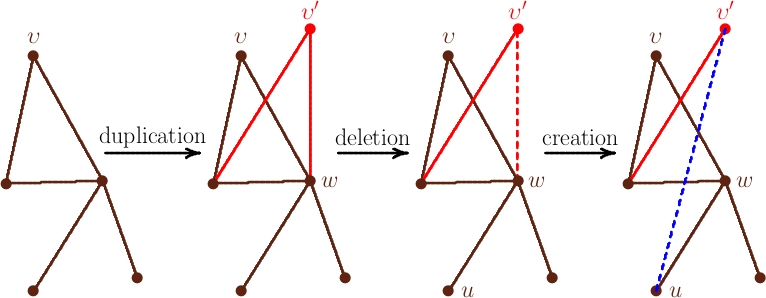}
  \caption{{The Sol\'e evolutionary algorithm:}
The duplication-deletion-creation iterations of the
Sol\'e algorithm.  A site is duplicated, some bonds incident on it are
deleted with probability $\delta$ and new bonds incident on it are
created with probability $\alpha$.}
\label{figure6S}
\end{figure}

The basic iterative step of the Sol\'e algorithm is shown in figure \ref{figure6S} and a
Sol\'e evolutionary network of order $N$ nodes is grown as follows:

\vspace{1mm}
\noindent{\bf Sol\'e evolutionary algorithm:}
\begin{enumerate}
\item[\bf 1.] Initiate the network with one node $x_0$ and apply the following steps iteratively;
\item[\bf 2.] Choose a node $\upsilon$ uniformly and duplicate it to a new node $\upsilon^\prime$;
\item[\bf 3.]  For each bond $\edge{w}{\upsilon}$ incident with the chosen node $\upsilon$, 
add the bond $\edge{w}{\upsilon^\prime}$ incident with the duplicated node
$\upsilon^\prime$;
\item[\bf 4.] Delete each bond $\edge{w}{\upsilon^\prime}$ added in step 3 with probability $\delta$
independently;
\item[\bf 5.] For all nodes $u$ not adjacent to the chosen node $\upsilon$, create the bond
$\edge{u}{\upsilon^\prime}$ with probability $\alpha$;
\item[\bf 6.] Stop the algorithm when a network of order $N$ is grown.
\qed
\end{enumerate}
\vspace{5mm}

The algorithm has two parameters $(\delta, \alpha)$.  If $\delta=0$ and $\alpha=1$
then the algorithm grows complete simple networks.  More generally, if $\alpha>0$ 
then on average roughly $\alpha N$ bonds are added to a network of order $N$.   This 
shows that the algorithm grows networks of size $O(N^2)$ -- that is, Sol\'e networks
are rich in bonds.

\subsection{Mean field theory for Sol\'e networks}

Let $E_n$ be the total number of bonds in a Sol\'e network after $n$ iterations 
of the algorithm, and let $\langle k \rangle_n$ be the connectivity of the network
(that is, the average degree of nodes) after $n$ iterations (so that 
$2\,E_n = n\langle k \rangle_n$).  In the mean field approximation  the node 
in step 2 of the algorithm has degree $\langle k \rangle_n$ and this number 
of bonds is added in step 3,  while, in a similar way, $\delta \langle k \rangle_n$ 
bonds are removed in step 4.  In step 5 there are $n-\langle k \rangle_n$ choices 
in the mean field for the node $u$ not adjacent to $\upsilon^\prime$ and 
each bond $\edge{u}{\upsilon^\prime}$ is added with probability $\alpha$.  
This shows that the number of bonds after $n+1$ iterations is given by
the recurrance relation
\begin{equation}
E_{n+1} = E_n + (1-\delta) \langle k \rangle_n + \alpha(n - \langle k \rangle_n) .
\end{equation}

Since $2\,E_n = n\langle k \rangle_n$ this becomes
\begin{equation}
E_{n+1} - E_n = \alpha n + \Sfrac{2}{n} (1-\delta-\alpha)\, E_n ,
\end{equation}
which is a mean field recurrance relation for $E_n$.

Taking $n\to t$, a continuous time variable, and approximating $E_n$ by $E_t$,
and approximating the finite difference as a derivative, gives the following
differential equation for $E_n$:
\begin{equation}
\sfrac{d}{dt} E_t = \alpha t + \Sfrac{2}{t}(1-\alpha-\delta)\, E_t .
\end{equation}
Solving this equation and letting $t\to n$ again gives the approximate mean
field solution for $E_n$:
\begin{equation}
E_n \approx \Sfrac{\alpha n^2}{2(\alpha+\delta)} 
+ \Sfrac{(\alpha+2\delta)n^{2(1-\alpha-\delta)}}{2(\alpha+\delta)} .
\label{eqn47a}  
\end{equation}
Equation \Ref{eqn47a} shows that the number of bonds is proportional to 
$n^2$, so that networks created by this algorithm are dense, except when
$\alpha=0$. Comparison to equation \Ref{eqn37} suggests that $\gamma\leq 1$ in this model.
Notice that there is no logarithmic factor in the denominator, and that
$E_n = \Theta( n^2 )$.  This is consistent with a mean field value $\gamma < 1$
(and this requires that $P_n(k)$ be modified so that it is a normalisable 
probability distribution).  With these results, it is reasonable to expect that,
in the mean field, 
\begin{equation}
\gamma \leq 1.
\label{eqn48a}   
\end{equation}
If $\alpha=0$ then equation \Ref{eqn47a} gives $E_n \sim n^{2-2\delta}$
and comparison to equation \Ref{eqn37} gives
\begin{equation}
\gamma = 1 + 2\delta, \quad\hbox{if $\alpha = 0$.}
\end{equation}

\begin{figure}[t!]
 \centering
\includegraphics[height=8cm]{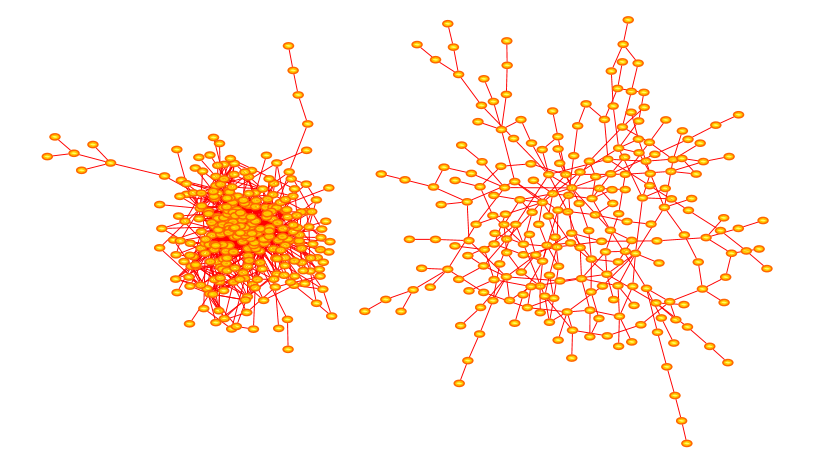}
\caption{{Sol\'e evolutionary networks:}
The network on the left was generated with $\delta=0.25$ and
$\alpha=0.005$.  Its has order $279$ and has $47$ nodes with
degrees exceeding $\sqrt{279}$ and so qualify as hubs.  The largest few of these
hubs have degrees $\{40,41,62,80\}$.  This algorithm creates dense networks
as seen here, even for small values of $\alpha$.  Increasing the value of
$\delta$ gives more extended networks.  The network on the right
was generated with $\delta=0.75$ and $\alpha=0.005$ and grown to order
$230$.  None of its nodes qualify as hubs.
 The arrangement of nodes and bonds in
these networks was created using the prefuse force directed lay-out in Cytoscape 3.4.0
\cite{Cytoscape}.}
\label{figure5S}
\end{figure}

\begin{figure}[t!]
 \centering
\includegraphics[height=8cm]{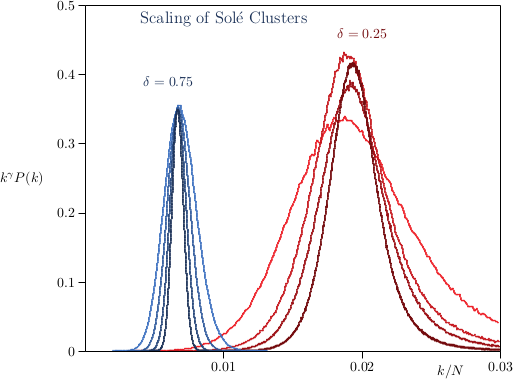}
\caption{{Scaling of Sol\'e evolutionary networks:}
Plotting $k^\gamma P_N(k)$ against $N^{-\phi}k$ for networks generated
by the Sol\'e Evolutionary algorithm gives the distributions above.
On the left the results are shown for networks grown with
$\delta=0.75$ and $\alpha=0.005$.  The choices $\gamma=1/2$ and
$\phi=1$ uncovers a distribution as shown where the order of the
networks are $N=100\times 2^n$ for $n=6,7,8,9$.  A similar
distribution, but with $\gamma=2/3$ and $\phi=1$, is seen
when networks are grown with $\delta=0.25$ and $\alpha=0.005$.
It is not known that the value of $\gamma$ changes discontinuously
as $\delta$ increases from $0.25$ to $0.75$.}
\label{figureSoleS}
\end{figure}

\subsection{Numerical results for Sol\'e networks}

Similar to Barabasi-Albert and Duplication-Divergence networks, Sol\'e networks
can be grown numerically by implementing the algorithm as given above, using
sparse matrix routines to efficiently store the adjacency matrix of the network.
The larger size of networks makes these more difficult to grow, and our 
algorithms sampled efficiently to networks of size $51200$ bonds.

Sol\'e networks are rich in bonds.  This is seen, for example, in equation
\Ref{eqn47a}, which shows that $E_n \propto n^2$ if $\alpha>0$.
In figure \ref{figure5S} two examples of networks generated by the Sol\'e 
algorithm are shown.  If $\delta < 0.5$, then the networks have a dense appearance 
dominated by a few hubs.  If $\delta > 0.5$, then the networks appear more extended, 
often with no nodes qualifying as hubs under the definition that the degree of a hub
in a network of order $n$ is at least $\lfloor \sqrt{n} \rfloor$.   The networks in 
figure \ref{figure5S} were generated with $\alpha=0.005$, and increasing the value 
of $\alpha$ quickly increases the number of bonds.

The mean field result that $\gamma \leq 1$ has implications for the scaling of Sol\'e 
networks.  In particular, $P_N(k)$ in equation \Ref{eqn1}  is not
normalisable for infinite networks if $\gamma \leq 1$ and so is not a valid
candidate degree distribution in this model.  
The degree distribution can be modified to
\begin{equation}
P (k) \simeq C_o\, k^{-\gamma} D(n^{-\phi} k)
\label{eqn52}  
\end{equation}
where $D(x)$ is a function of the combined (or scaled) variable $x=n^{-\phi}k$.  
That is, as $n\to\infty$, $k$ is rescaled by $n^{-\phi}$
and $k^\gamma P(k)$ approaches a limiting distribution proportional to
$D(x)$.  

This can be tested numerically by plotting $n^\gamma P(k)$ as a function 
of $x=n^{-\phi}k$.  For the proper choices of $\gamma$ and $\phi$
it is expected that $n^\gamma P(k) \simeq C_o D(x)$ for a wide range 
of values of $n$ (that is, the data should approach a limiting curve as $n\to\infty$).  
The result is shown in figure \ref{figureSoleS} for $(\delta=0.25,\alpha=0.005)$ 
and $(\delta=0.75,\alpha=0.005)$. These are plots on the same graph for 
$n = 100 \times 2^n$ for $n\in\{6,7,8,9\}$ (other curves at smaller values of $N$
are left away to give a clearer picture).

The data for $\delta=0.75$ are the cluster of peaks to the left, rescaled by
choosing $\phi=1$ and $\gamma=\sfrac{1}{2}$, while the cluster of peaks to the
right is for $\delta=0.25$ with $\phi=1$ and $\gamma=\sfrac{2}{3}$.  With increasing
$n$ the data appear to approach a single underlying curve if $\gamma=\sfrac{1}{2}$ in the
one instance, and $\gamma=\sfrac{2}{3}$ in the other instance.  Both these
values are consistent with the mean field expectation that $\gamma \leq 1$ in
this model.  Further refinements in this scaling assumption may be necessary, since
the curves are still becoming narrower with increasing $n$.  It is not clear that
these approach a limiting curve as $n\to\infty$, although the data for $\delta=0.75$
suggest this to be the case.   In these cases the curves are sharply peaked 
with a mean of about $0.02$ if $\delta=0.25$ and about $0.007$ if $\delta=0.75$.

\begin{table}[t!]
\centering
\caption{Connectivity data for Sol\'e Networks.}
      \begin{tabular}{lccc}
        \hline
       $n$    &  $\delta=0.25$  & $\delta = 0.75$\\ \hline
        $100$ & $2.95$ & $1.50$ \\
        $200$ & $4.46$ & $1.94$ \\
        $400$ & $7.59$ & $2.94$ \\
        $800$ & $14.75$ & $5.36$ \\
        $1600$ & $30.46$ & $10.64$ \\
        $3200$ & $59.94$ & $21.26$ \\
        $6400$ & $122.78$ & $45.57$ \\
        $12800$ & $245.35$ & $85.18$ \\
      $25600$ & $496.87$ & $170.35$ \\
      $51200$ & $994.54$ & $340.76$ \\ \hline
      \end{tabular}
\label{tableSole}
\end{table}

Since the curve $D(x)$ is sharply peaked at a constant value $c_o$ of the
rescaled variable $x$, the connectivity of Sol\'e networks is estimated by treating
$D(x)$ as concentrated at $c_o$ and then (assuming that $\phi=1$ and approximating
the connectivity)
\begin{equation}
\langle k \rangle_n \sim
\frac{\int_0^\infty k^{1-\gamma} D(k/n^\phi)\, dk}{\int_0^\infty k^{-\gamma} D(k/n^\phi)\, dk}
\sim \frac{(n^{\phi})^{2-\gamma}}{(n^\phi)^{1-\gamma}} \sim n^\phi .
\end{equation}
In other words, the connectivity of Sol\'e networks should increase linearly with
$n^\phi$ (and since $\phi=1$, linearly with $n$).  In table \ref{tableSole} the 
connectivities of Sol\'e networks for $\delta=0.25$ and $\delta=0.75$
(with $\alpha=0.005$) are listed.  Non-linear least squares fits to the data show that
$\phi=1.01$ when $\delta=0.25$ and $\phi=0.99$ when $\delta=0.75$.  That is, these
results are consistent with the value $\phi=1$ seen above.

\section{The iSite model of network evolution}

Protein interaction networks evolve by mutations in proteins which change the interactions of
the proteins in the network.  In the Duplication-Divergence algorithm, a mutated protein loses its
interactions randomly.   This random deletion of interactions is a good first order approximation
to the evolution of networks.  The iSite model refines this by giving structure to nodes in the
network by introducing \textit{iSites} on nodes as localities of the interaction sites on a protein
\cite{G11,GG15}. Subfunctionalization of interaction sites in the iSite model is implemented by
silencing iSites, and adding interactions with reduced probability if the iSite is not silenced.

The implementation of the iSite algorithm relies in the first place on duplication of nodes, and
then subfunctionalization of iSites on the nodes.   The subfunctionalization of iSites is implemented
by randomly deleting of bonds incident to duplicated iSites, \textit{and} by the silencing
of iSites by turning them off.  These processes are models of random mutations which cause
the loss of information in the genome (and leave behind non-coding remnants of genes).
A process of spontaneously creating new iSites is not in the iSites algorithm, although this is
a possible refinement which may be introduced in the algorithm.

\begin{figure}[t!]
 \centering
\includegraphics[height=2.75in]{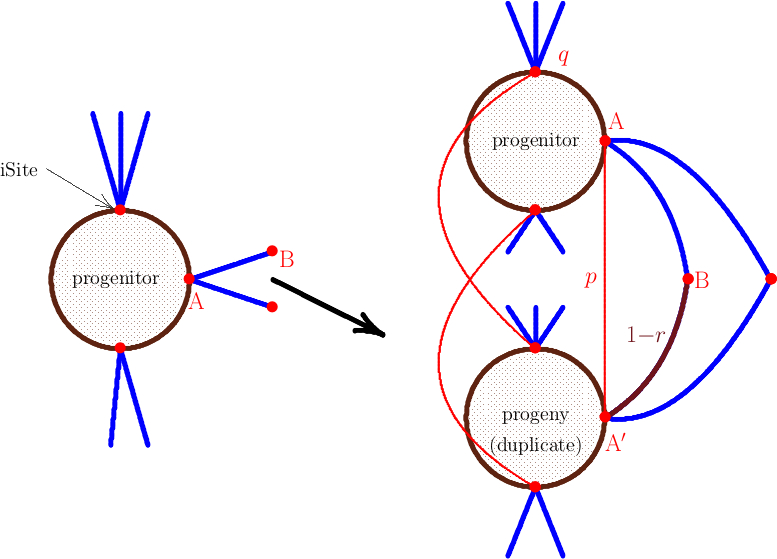}
  \caption{{The iSite evolutionary algorithm:}
The duplication-deletion iterations of the iSite algorithm.  A node together
with its iSites is duplicated, and some bonds incident with the duplicated
iSites are deleted with probability $r$.  New bonds between a self-interacting
iSite and its duplicate are inserted with probability $p$, and iSites are
silenced with probability $q$.}
\label{figureiSitealgorithm}
\end{figure}

The elementary move of the iSite algorithm is illustrated schematically in figure
\ref{figureiSitealgorithm}.  A uniformly chosen node is duplicated into a progenitor-progeny 
pair (and so also duplicating the iSites of the progenitor onto the progeny).   
If the duplicated iSite is self-interacting, then bonds are added between the iSite on the 
progenitor and the duplicated iSite on the progeny with probability $p$ -- this allows
for subfunctionalization of the duplicate iSites.  Bonds incident with the iSites on the
progenitor are duplicated with reduced probability $r$, and iSites on the progenitor or 
progeny nodes are silenced with probability $q$.  If an iSite is silenced, then all bonds
incident with it are deleted.  Notice that subfunctionalization enters in several ways,
both in the duplication of self-interacting iSites, in the duplication of bonds, and in
the silencing of iSites.

\vspace{1mm}
\noindent{\bf iSite evolutionary algorithm:} 
\begin{enumerate}
\item[\bf 1.] Initiate the network with one node $x_0$ with $I$ active iSites (each of which is
self-interacting with probability $p$) and iterate the following steps;
\item[\bf 2.] Choose a progenitor protein $\upsilon$ uniformly in the network and duplicate it, and its
associated iSites $A$, to a successor protein $\upsilon^\prime$ with duplicated iSites $A^\prime$;
\begin{itemize}
\item[\bf (a)] A duplicated iSite $A^\prime\in\upsilon^\prime$ is active with probability $q$
if it is duplicated from an active iSite on $A\in\upsilon$, and silenced otherwise;
\item[\bf (b)] An active duplicated iSite $A^\prime\in\upsilon^\prime$ is self-interacting with probability
$p$ if it is duplicated from a self-interacting iSite on $A\in\upsilon$, and not self-interacting
otherwise;
\item[\bf (c)] If a silenced iSite $A$ is duplicated to iSite $A^\prime$, then $A^\prime$ is also silenced;
\end{itemize}
\item[\bf 3.] Add bonds as follows:
\begin{itemize}
\item[\bf (a)] If iSite $A\in\upsilon$ is self-interacting and $A$ is duplicated to iSite 
$A^\prime\in\upsilon^\prime$, then add the bond $\edge{A}{A^\prime}$ 
if $A^\prime$ is not silenced;
\item[\bf (b)] If $\edge{A}{B}$ is a bond incident with iSite $A$ on the progenitor $\upsilon$, and
$A$ is duplicated to iSite $A^\prime$ on the duplicate $\upsilon^\prime$, then
$\edge{A}{B}$ is duplicated to $\edge{A^\prime}{B}$ with probability $1-r$ 
provided that $A^\prime$ is not silenced;
\end{itemize}
\item[\bf 4.] Iterate the algorithm from step (2) and stop the iterations when a network of order $N$ is grown.
\qed
\end{enumerate}
\vspace{5mm}

\subsection{Mean field theory for the iSite model}
Let nodes in the network correspond to proteins, and let $i_j(n)$ be the number of active
iSites on node $j$ after $n$ iterations of the algorithm.  Denote the degree of node $j$ 
by $k_j(n)$ (that is the total number of bonds with one end-point in node $j$),
and let $E_n$ be the number of bonds of the network (this is the \textit{size} of
the network).  Then $2\thin E_n = \sum_j k_j(n)$.

The average number of active iSites per node is $i(n) = \sfrac{1}{n} \sum_j i_j(n)$.  With each iteration
$i(n)$ iSites are created, of which $q\thin i(n)$ are silenced, in the mean field.  This gives the following
recurrance relation for $i(n)$:
\begin{equation}
(n+1)\,i(n+1) = n\, i(n) + (1-q)\,i(n) .
\label{eqn44x}   
\end{equation}
The exact solution of this recurrance is
\begin{equation}
i(n) = \Sfrac{i(0)\, \Gamma(1-q+n)}{n!\,\Gamma(1-q)}
\label{eqn45}   
\end{equation}
where $\Gamma$ is the gamma function with the property that $\Gamma(x+1)=x\,\Gamma(x)$ and
$\Gamma(1)=1$.   Notice that $i(0) = I$, where $I$ is the number of iSites on the source node
$x_0$.

For large $n$ the $\Gamma$-function and the factorial have well known asymptotics
(namely the Stirling approximation \cite{W61}), so that
\begin{equation}
i(n) \simeq \Sfrac{I\,n^{-q}}{\Gamma(1-q)} .
\label{eqn46}   
\end{equation}
This shows that with increasing $n$ the total number of iSites grows proportionally to $n^{1-q}$.
If $q=0$, then this is linear in $n$ since no iSites become silenced, and if $q=1$, then the 
number approaches a constant.

The total number of bonds in the network increases after $n$ iterations by the recurrance
\begin{equation}
E_{n+1} = E_n + \Sfrac{2(1-r)}{n}\,E_n + p\, i(n) ,
\label{eqn47}   
\end{equation}
since there are on average $\sfrac{2}{n} E_n$ bonds incident to each node, and the probability
that each one of them is duplicated is $1-r$, and there are on average $i(n)$ iSites
per node, and the probability that each of these is self-interacting is $p$.

Using the asymptotic solution for $i(n)$ and approximating this recurrence by a 
differential equation gives
\begin{equation}
\sfrac{d}{dt}\,E_t = \Sfrac{2(1-r)}{t}\,E_t + \Sfrac{pI}{\Gamma(1-q)}\,t^{-q} .
\end{equation}
This equation can be solved, and using the initial condition $E_1=0$, the result is
\begin{equation}
E_t = \Sfrac{pI}{(1+q-2r)\,\Gamma(1-q)} \left( t^{2-2r} - t^{1-q} \right).
\label{eqn49}   
\end{equation}
Thus, the average degree of a node is equal to $\sfrac{2}{n} E_n$, so that the
connectivity of iSite evolutionary networks is given by
\begin{equation}
\langle k\rangle_n \simeq \Sfrac{2pI}{(1+q-2r)\,\Gamma(1-q)} \left( t^{1-2r} - t^{-q} \right) 
\label{eqn50}   
\end{equation}
in the mean field. This shows that the large $n$ value of $\langle k \rangle_n$ 
is dominated by the larger of $-q$ and $1-2r$.  In particular, if $r < \sfrac{1}{2}(1+q)$, 
then $\langle k \rangle_n \sim n^{1-2r}$.

By equation \Ref{eqn37} one may  determine $\gamma$ for this model: 
\begin{equation}
\gamma = 
\cases{
 1+2r, &\hbox{if $r<\sfrac{1}{2}(1+q)$}; \\
 2+q, &\hbox{if $r>\sfrac{1}{2}(1+q)$}.
}
\label{eqn51}   
\end{equation}
If $2r=1+q$, then a different solution is obtained, namely
\begin{equation}
E_t = \Sfrac{pI}{\Gamma(1-q)}\, t^{1-q}\log t .
\label{eqn52x}   
\end{equation}
This shows that $\gamma=2+q$ in this case as well, but there is also a logarithmic
correction to the growth of $E(t)$, and so there is a logarithmic factor in
the expression for $\langle k \rangle_n$.

\begin{figure}[t!]
 \centering
\includegraphics[height=9cm]{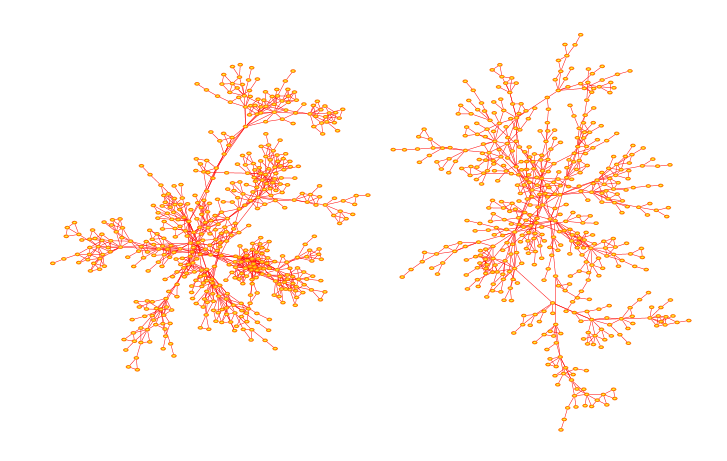}
\caption{{iSite evolutionary networks:}
The network on the left was generated with $4$ iSites per node,
$p=0.5$, $q=0.1$ and $r=0.8$, and the network on the right was generated with
$2$ iSites per node, and with $p=0.5$, $q=0.1$ and $r=0.8$.
The order of the network on the left is $501$ and on the right, $491$.
The network on the left has two nodes qualifying as hubs, of degrees
$\{23,25\}$, while the network on the right has none.
 The arrangement of nodes and bonds in
these networks was created using the prefuse force directed lay-out in Cytoscape 3.4.0
\cite{Cytoscape}.}
\label{figure6M}
\end{figure}

\subsection{Modified iSite evolutionary algorithm}

The subfunctionalization of proteins can be refined by introducing in the
iSite algorithm the probability of creating new iSites on the progeny node
with a probability $s$.  This changes the algorithm as follows.

\vspace{1mm}
\noindent{\bf Modified iSite evolutionary algorithm:}  
Implement the algorithm as above but introduce the parameter $s$ and create 
new active iSites by replacing step 2 in the iSite evolutionary algorithm by 
\begin{itemize}
\item[\bf 2.] Choose a progenitor node $\upsilon$ uniformly in the network and duplicate it, and its
associated iSites $A$, to a progeny node $\upsilon^\prime$ with duplicated iSites $A^\prime$;
\begin{itemize}
\item[\bf (a)] A duplicated iSite $A^\prime\in\upsilon^\prime$ is active with probability $q$
if it is duplicated from an active iSite on $A\in\upsilon$, and silenced otherwise;
\item[\bf (b)] An active duplicated iSite $A^\prime\in\upsilon^\prime$ is self-interacting with probability
$p$ if it is duplicated from a self-interacting iSite on $A\in\upsilon$, and not self-interacting
otherwise;
\item[\bf (c)] If a silenced iSite $A$ is duplicated to iSite $A^\prime$, then $A^\prime$ is also silenced;
\item[\bf (d)] With probability $s$ create an active iSite $C$ on the progeny 
node $\upsilon^\prime$, where $C$ is self-interacting with probability $p$.
\end{itemize}
\end{itemize}

The recurrance for the average number of active iSites per node $i(n)$ (see equation
\Ref{eqn45}) is modified to
\begin{equation}
(n+1)\,i(n+1) = n\,i(n) + (1+s-q)\, i(n) 
\end{equation} 
in the Modified iSite evolutionary algorithm.  The exact solution is obtained by replacing 
$q$ by $q-s$ in equation \Ref{eqn45}, and the asymptotic approximation of the
solution is given by
\begin{equation}
i(n) \simeq \Sfrac{I\, n^{s-q}}{\Gamma(1+s-q)} ,
\end{equation}
as seen in equation \Ref{eqn46}.

The total number of bonds in the network, $E_n$, still satisfies equation \Ref{eqn47},
and so it follows from equations \Ref{eqn49}, \Ref{eqn50} and \Ref{eqn51},
that for the modified iSite evolutionary algorithm (notice the condition that $q<r+s$):
\begin{equation}
E_n = \Sfrac{pI}{(1+q-s-2r)\,\Gamma(1+s-q)} \left( n^{2-2r} - n^{1+s-q} \right) .
\end{equation}
This shows that the connectivity of Modified iSite networks is given by 
\begin{equation}
\langle k\rangle_n \simeq \Sfrac{2pI}{(1+q-s-2r)\,\Gamma(1+s-q)} \left( n^{1-2r} - n^{s-q} \right) .
\end{equation}
The value of the scaling exponent is seen from above to be given by
\begin{equation}
\gamma = 
\cases{
 1+2r, & \hbox{if $r<\sfrac{1}{2}(1+q-s)$};  \\
 2+q-s, & \hbox{if $r>\sfrac{1}{2}(1+q-s)$}.
}
\label{eqn51M}   
\end{equation}
with a correction factor in the expression for $\langle k \rangle_n$ if
$2r=(1+q-s)$.

\subsection{Numerical results for iSite networks}

The iSite algorithm was coded and networks were grown to compute
averaged statistics.  Examples of iSite networks generated by the 
algorithm are shown in figure \ref{figure6M}.  The algorithm was then
used to sample networks of size up to $200,000$.

The connectivity $\langle k \rangle_n$ of iSite networks for $I=3$ iSites per node, 
and with $p=0.5$, $q=0.4$ and $r=0.3$, is shown in table \ref{tableiSite}.  
By equation \Ref{eqn3}, $\log \langle k \rangle \simeq \log \sfrac{\gamma-1}{2-\gamma}
+ (2-\gamma) \log n$.  Least squares fit to the data in Column 2 gives
$\log \sfrac{\gamma-1}{2-\gamma} \approx 1.0211$, and 
$(2-\gamma) = 0.258$.  Solving for $\gamma$ gives in the first instance
$\gamma = 1.735$ and in the second $\gamma=1.742$.  Since
$2r < 1+q$ in this case, the mean field value of $\gamma$ is $\gamma=1+2r = 1.6$,
close to these estimated values.

\begin{table}[t!]
\centering
\caption{Connectivity data for iSite Networks.}
      \begin{tabular}{lccccc}
        \hline
       $n$    &  Column 2  & Column 3 & Column 4 & Column 5 \\ \hline
        $3125$ & $22.385$ & $20.701$ & $4.756$ & $6.648$  \\
        $6250$ & $26.524$ & $25.752$ & $4.770$ & $6.556$ \\
        $12500$ & $31.395$ & $29.137$ & $4.677$ & $6.579$ \\
        $25000$ & $37.808$ & $35.308$ & $4.733$ & $6.358$ \\
        $50000$ & $45.931$ & $42.244$ & $4.579$ & $6.299$ \\
        $100000$ & $54.830$ & $50.035$ & $4.584$ & $6.204$ \\
        $200000$ & $64.668$ & $59.284$ & $4.649$ & $6.071$ \\ \hline
   Column 2: & $I=3$,\;$p=0.5$ & $q=0.4$,\;$r=0.3$ \\ 
   Column 3: & $I=5$,\;$p=0.5$ & $q=0.4$,\;$r=0.3$ \\
   Column 4: & $I=3$,\;$p=0.5$ & $q=0.05$,\;$r=0.8$ \\ 
   Column 5: & $I=5$,\;$p=0.5$ & $q=0.05$,\;$r=0.8$ \\ \hline
      \end{tabular}
\label{tableiSite}
\end{table}

Data for $I=5$ and with the same values of $(p,q,r)=(0.5,0.4,0.3)$ are shown in
table \ref{tableiSite} as well.  Changing the value of $I$ (the number of iSites per
node) should not change the value of $\gamma$, and this appears to be the
case here.  A least squares fit to the data in Column 3 and determining 
$\gamma$ as above gives $\gamma=1.737$ and $\gamma=0.7498$, very close
to the values above.

If $p=0.5$, $q=0.05$ and $r=0.8$, then $2r > 1+q$, and in this case $\gamma=2+q$.
If the number of iSites per node is $I=3$, then the data in table \ref{tableiSite} 
gives a constant value for $\langle k \rangle$, and for $I=5$ a slightly decreasing 
numerical estimate.  The mean field value of $\gamma$ in these cases is $2.05$, 
and a least squares fit gives $\gamma\approx 2.009$ if $I=3$ and 
$\gamma\approx 2.022$ if $I=5$ (where the coefficient of $\log n$ in the least 
squares fit is $2-\gamma$).  These results are consistent with the mean field 
results obtained above, since it shows that the value of $\gamma$ is close to $2+q$.

\begin{figure}[t!]
 \centering
\includegraphics[height=8cm]{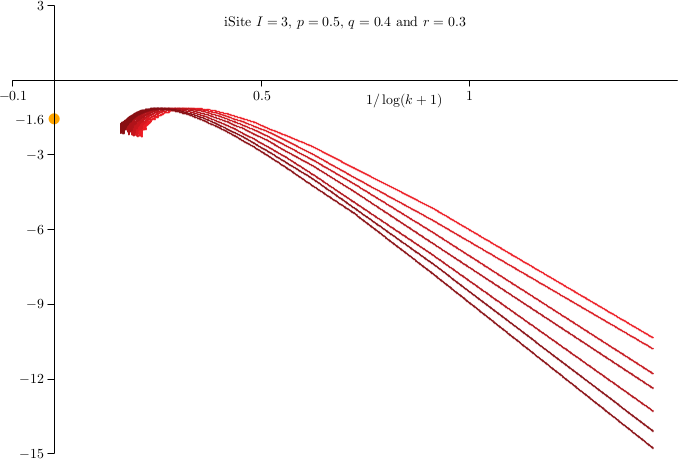}
\caption{{iSite evolutionary networks with $I=3$, $p=0.5$,
$q=0.4$ and $r=0.3$:}  
Data on networks generated by the iSite evolutionary algorithm.  In each case $500$ 
networks were grown and the average degree sequence $P_n(k)$ computed.  The curves 
are plots of $\log P_n(k)/\log (k+1)$ against $1/\log (k+1)$ for $n\in\{3125,6250,12500,\cdots,200000\}$.
As $k\to\infty$, then the curves are expected to pass through $-\gamma$ on the
$y$-axis, and its mean field value is $\gamma = 1+2r=1.6$ -- this value is marked on the
$y$-axis.}
\label{figure6T}
\end{figure}

\section{Conclusions}

In this paper a number of algorithms used for generating networks in molecular biology
were examined.  Mean field theory for the algorithms was in some cases reviewed,
and in other cases newly presented, and also refined.  The algorithms include the
Barabasi-Albert \cite{AB02}, Duplication-Divergence \cite{TR04}, Sol\'e \cite{SPSK02}
and iSite algorithms \cite{G11,GG15}, and these were in
some cases modified by the introduction of more general elementary moves.

The efficient implementation of these algorithms was also examined, and sparse
matrix routines (or, more general, hash-coding; see for example reference \cite{Knuth98})
were used to  optimize the implementation.  This gives computer algorithms which
can generate very large networks efficiently, and networks of order $200,000$ nodes
were routinely sampled.  We also explored even larger networks, up to order 3 million,
but did not use those in our data analysis.  

The adjacency matrix of a network of size $E$ bonds can be stored (using sparse 
matrix routines) in an array of size $O(E)$.  This means that the implementation
of these network growth algorithms has average case space complexity
$O(E)$.

Hash coding allows for the efficient implementation of routines which search,
insert or delete entries in arrays storing the networks.  These routines have
average time complexity $O(1)$ \cite{Cormen09}, (and worst case time complexity
$O(E)$ for searches, inserting and deleting bonds, due to collisions if a hash
table is densely populated).

Generally, the time complexity of algorithms should grow
as $O(E^\tau)$ if networks of size $E$ are grown (where $\tau$ is an exponent
dependent on the particular algorithm).   For example, networks of size $E$ 
bonds can be generated using $O(E)$ computer memory, and the 
Duplication-Divergence and iSite algorithms can be implemented with
$O(n^\tau)$ time complexity to grow networks of order $n$ nodes (and where
$n\leq E$).  An examination of these algorithms (the Duplication-Divergence
and iSite algorithms) suggests that an optimal implementation will have
$\tau\approx 1$ (if the size of the hash tables is much larger than $n$).  

The Barabasi-Albert and Sol\'e algorithms (with their modified and variant 
implementations) should have average time complexity of $O(n^2)$ for
growing networks of order $n$ nodes.  This follows because each iteration of the
algorithms has to explore all nodes in the current network for the possible
insertion of new bonds.

\begin{table}[t!]
\caption{Computational Time Complexity of Implemented Algorithms.}
\scalebox{0.75}{
      \begin{tabular}{lcccc|r}
        \hline
         Algorithm  & $n=6250$  & $n=12500$  & $n=25000$ & $n=50000$ & $\tau$ \\ \hline
        Bar-Alb    ($p=0$)                & $0.602$ & $2.51$ & $9.03$ & $38.0$ & $1.97$ \\
        Mod Bar-Alb ($\lambda=2$, $p=A=0$) & $0.618$ & $2.55$ & $10.1$ & $36.3$ & $1.96$ \\
        Var Bar-Alb ($\alpha=2$, $a=0$) & $1.35$ & $4.46$ & $16.4$ & $--$ & $--$ \\
        Dupl-Div  ($p=1$, $q=0.4$) & $0.349$ & $0.862$ & $2.04$ & $5.01$ & $1.28$ \\ 
        Dupl-Div  ($p=1$, $q=0.6$) & $0.155$ & $0.319$ & $0.635$ & $1.31$ & $1.02$ \\ 
        Mod Dupl-Div  ($p=1$, $q=0.4$) & $0.340$ & $0.891$ & $2.45$ & $7.09$ & $1.46$ \\ 
        Mod Dupl-Div  ($p=1$, $q=0.6$) & $0.165$ & $0.338$ & $0.699$ & $1.44$ & $1.04$ \\ 
        Sol\'e      ($\delta=0.25$, $\alpha=0.005$)              & $4.84$ & $20.5$ & $91.0$ & $436.0$ & $2.16$ \\ 
        Sol\'e      ($\delta=0.75$, $\alpha=0.005$)              & $6.10$ & $20.0$ & $79.5$ & $323.2$ & $1.92$ \\ 
        iSite        ($p=0.5$, $q= 0.01$, $r=0.8$, $I=1$)       & $0.114$ & $0.234$ & $0.454$ & $0.925$ & $1.00$ \\ 
        iSite        ($p=0.5$, $q= 0.01$, $r=0.8$, $I=2$)       & $0.110$ & $0.216$ & $0.458$ & $0.878$ & $1.01$ \\ 
        iSite        ($p=0.5$, $q= 0.01$, $r=0.8$, $I=3$)       & $0.106$ & $0.217$ & $0.432$ & $0.857$ & $1.00$ \\ 
        iSite        ($p=0.5$, $q= 0.01$, $r=0.8$, $I=4$)       & $0.107$ & $0.231$ & $0.422$ & $0.848$ & $0.98$ \\ 
        iSite        ($p=0.25$, $q= 0.01$, $r=0.8$, $I=4$)       & $0.104$ & $0.249$ & $0.415$ & $0.844$ & $0.98$ \\ 
        iSite        ($p=0.75$, $q= 0.01$, $r=0.8$, $I=4$)       & $0.108$ & $0.216$ & $0.437$ & $0.867$ & $1.00$ \\ 
        Mod iSite   ($p=0.5$, $q= 0.1$, $r=0.8$, $s=0.1$, $I=4$)        & $0.288$ & $0.560$ & $1.102$ & $2.53$ & $1.04$ \\ 
     \hline
      \end{tabular}
}
\label{complexity}
\end{table}

Data on the time complexity of the algorithms are shown in table \ref{complexity}.  The data
displayed are the average time $T$ to grow one network of order $n$.  Assuming
that $T=C_0 n^\tau$ and fitting $\log T$ to $\log n$, least squares estimates of
$\tau$ can be obtained.  For example, it is expected that $\tau=2$ for the Barabasi-Albert
algorithm, while the estimate obtained in the table is $\tau\approx 1.97$.  This is
consistent with the expectation that the time complexity of the algorithm is $O(n^2)$
in an optimal implementation.  This is similarly seen for the modified and variant
implementation of the Barabasi-Albert algorithm, and for the Sol\'e algorithm.

The time complexity of the remaining algorithms is $O(n)$, and this is found consistently,
except for the Duplication-Divergence algorithm for $q=1$ and $q=0.4$ (and also
for the modified implementation of this algorithm).  In these cases the algorithm
samples denser networks (see figure \ref{figure4}) which takes up larger amounts of
memory, making the implementation less efficient.

\begin{figure}[t!]
 \centering
\includegraphics[height=8cm]{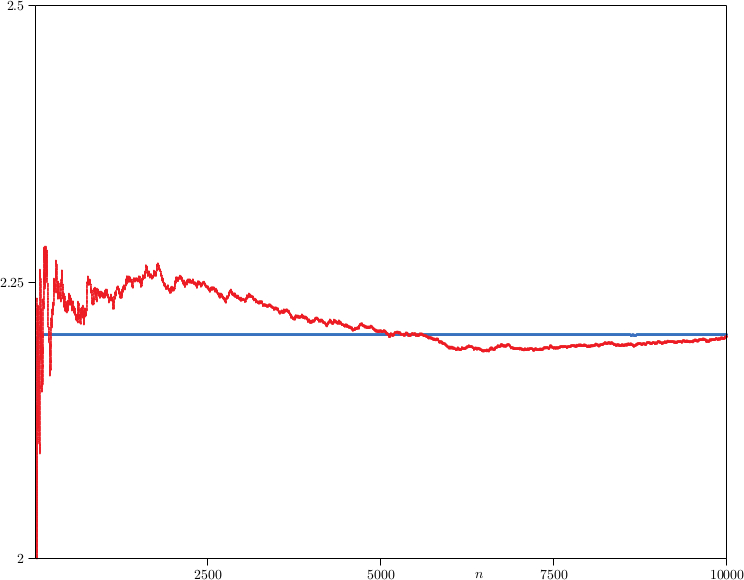}
\caption{{Self-averaging of the connectivity of Barabasi-Albert networks:}
The connectivity of a single network grown with the Barabasi-Albert algorithm with
$p=0.6$ as a function of the size of the network is given by the noisy red curve as
the network is grown to order $n=10000$.  The blue curve is the average connectivity
of Barabasi-Albert networks, plotted as a function of $n$.  Notice that the red data
appear to converge, with increasing $n$ to the average, so that the connectivity of 
a randomly grown Barabasi-Albert network appears to converge to its average.}
\label{figure16}
\end{figure}

The results in this paper raise some questions about the sampling of scale-free networks by random iterative growth algorithms: 
\begin{itemize} 
\item In some cases, see for example reference \cite{VF02}, the parameters of the algorithms 
were set to grow networks with properties similar to that of real protein 
interaction networks. The values of the parameters are then used to estimate 
the rate of subfunctionalization (or mutation) in the genome. The results are dependent on 
the algorithm, and so further refinement of algorithms may be needed before useful estimates 
can be made. 
\item The mean field approaches are useful in some models (for example the 
Barabasi-Albert algorithm, and the iSite algorithm), but are poorer approximations in 
other models (the variant Barabasi-Albert algorithm, the Duplication-Divergence 
algorithm and its modification, and the Sol\'e algorithm). Can the mean field approach 
be improved to give a better approximation to these algorithms? 
\item Investigation of some numerical properties of the networks (for example the connectivity) 
suggests that the algorithms may be self-averaging. That is, networks are generated 
with properties which converge to the statistical averages of these properties 
over a sample of networks generated by the algorithm.  This is, for example, illustrated
in figure \ref{figure16} for the connectivity of Barabasi-Albert networks.  As the network is
grown, its connectivity appears to approach the average connectivity over a large sample
of networks. 
\item In this paper some algorithms were modified in ways not done before in 
the literature (this includes the modified Barabasi-Albert, the Duplication-Divergence, 
the Sol\'e and iSite models).  Exploring the properties of these modified algorithms,
including their usefulness as models of networks in molecular biology, will be the
subject of future investigation.
\end{itemize} 

Lastly, these algorithms grow networks using a probabilistic set of rules to
implement an elementary move.  Each realised network $N_n$ of order $n$ 
is obtained with some probability $p(N_n)$, so that the function $p(N_n)$ is 
a probability distribution over networks of order $n$.  Determining $p(N_n)$ for
any of the algorithms presented here seems difficult, and general properties
of $p(N_n)$ remain unknown (other than averages of network properties
over $p(N_n)$ are scale-free if the algorithm grows scale-free networks).

\section*{Acknowledgements}
  EJJvR is grateful to NSERC (Canada) for support in the form of a Discovery Grant

\section*{References}
\bibliographystyle{plain} 
\bibliography{scalefree}      

\begin{thebibliography}{10}

\bibitem{Cytoscape}
{\em Cytoscape 3.4.0}.
\newblock Cytoscape Developers, NRNB, 2016.

\bibitem{AB02}
R~Albert and A-L Barab{\'a}si.
\newblock Statistical mechanics of complex networks.
\newblock {\em Reviews of Modern Physics}, 74(1):47--97, 2002.

\bibitem{B09}
A-L Barab{\'a}si.
\newblock Scale-free networks: a decade and beyond.
\newblock {\em science}, 325(5939):412--413, 2009.

\bibitem{BA99}
A-L Barab{\'a}si and R~Albert.
\newblock Emergence of scaling in random networks.
\newblock {\em Science}, 286(5439):509--512, 1999.

\bibitem{BAJ99}
A-L Barab{\'a}si, R~Albert, and H~Jeong.
\newblock Mean-field theory for scale-free random networks.
\newblock {\em Physica A: Statistical Mechanics and its Applications},
  272(1):173--187, 1999.

\bibitem{BAJ00}
A-L Barab{\'a}si, R~Albert, and H~Jeong.
\newblock Scale-free characteristics of random networks: the topology of the
  world-wide web.
\newblock {\em Physica A: Statistical Mechanics and its Applications},
  281(1):69--77, 2000.

\bibitem{BBP04}
A~Barrat, M~Barthelemy, R~Pastor-Satorras, and A~Vespignani.
\newblock The architecture of complex weighted networks.
\newblock {\em Proceedings of the National Academy of Sciences of the United
  States of America}, 101(11):3747--3752, 2004.

\bibitem{BLW04}
J~Berg, M~L{\"a}ssig, and A~Wagner.
\newblock Structure and evolution of protein interaction networks: a
  statistical model for link dynamics and gene duplications.
\newblock {\em BMC evolutionary biology}, 4(1):1--12, 2004.

\bibitem{BLM06}
S~Boccaletti, V~Latora, Y~Moreno, M~Chavez, and D-U Hwang.
\newblock Complex networks: Structure and dynamics.
\newblock {\em Physics reports}, 424(4):175--308, 2006.

\bibitem{Cormen09}
TH~Cormen.
\newblock {\em Introduction to algorithms}.
\newblock MIT {P}ress, Cambridge {MA}, 2009.

\bibitem{EL03}
Eli Eisenberg and Erez~Y Levanon.
\newblock Preferential attachment in the protein network evolution.
\newblock {\em Physical review letters}, 91(13):138701, 2003.

\bibitem{E06}
E~Estrada.
\newblock Virtual identification of essential proteins within the protein
  interaction network of yeast.
\newblock {\em Proteomics}, 6(1):35--40, 2006.

\bibitem{EI08}
K~Evlampiev and H~Isambert.
\newblock Conservation and topology of protein interaction networks under
  duplication-divergence evolution.
\newblock {\em Proceedings of the National Academy of Sciences},
  105(29):9863--9868, 2008.

\bibitem{G11}
T~A Gibson and D~S Goldberg.
\newblock Improving evolutionary models of protein interaction networks.
\newblock {\em Bioinformatics}, 27(3):376--382, 2011.

\bibitem{GG15}
T~A Gibson and D~S Goldberg.
\newblock The topological profile of a model of protein network evolution can
  direct model improvement.
\newblock In {\em International Workshop on Algorithms in Bioinformatics},
  pages 40--52. Springer, 2015.

\bibitem{GA05}
R~Guimera and L~A~N Amaral.
\newblock Functional cartography of complex metabolic networks.
\newblock {\em Nature}, 433(7028):895--900, 2005.

\bibitem{HZ05}
Xionglei He and Jianzhi Zhang.
\newblock Rapid subfunctionalization accompanied by prolonged and substantial
  neofunctionalization in duplicate gene evolution.
\newblock {\em Genetics}, 169(2):1157--1164, 2005.

\bibitem{JTA00}
H~Jeong, B~Tombor, R~Albert, Z~N Oltvai, and A-L Barab{\'a}si.
\newblock The large-scale organization of metabolic networks.
\newblock {\em Nature}, 407(6804):651--654, 2000.

\bibitem{KBL04}
M~Kellis, B~W Birren, and E~S Lander.
\newblock Proof and evolutionary analysis of ancient genome duplication in the
  yeast saccharomyces cerevisiae.
\newblock {\em Nature}, 428(6983):617--624, 2004.

\bibitem{KW06}
Raya Khanin and Ernst Wit.
\newblock How scale-free are biological networks.
\newblock {\em Journal of computational biology}, 13(3):810--818, 2006.

\bibitem{Knuth98}
DE~Knuth.
\newblock {\em The art of computer programming: {S}orting and searching},
  volume~3.
\newblock Addison-Wesley, Reading MA, 1973.

\bibitem{NSRA12}
J~N{\"a}svall, L~Sun, J~R Roth, and D~I Andersson.
\newblock Real-time evolution of new genes by innovation, amplification, and
  divergence.
\newblock {\em Science}, 338(6105):384--387, 2012.

\bibitem{PS03}
R~Pastor-Satorras, E~Smith, and R~V Sol{\'e}.
\newblock Evolving protein interaction networks through gene duplication.
\newblock {\em Journal of Theoretical biology}, 222(2):199--210, 2003.

\bibitem{SPSK02}
R~V Sol{\'e}, R~Pastor-Satorras, E~Smith, and T~B Kepler.
\newblock A model of large-scale proteome evolution.
\newblock {\em Advances in Complex Systems}, 5(01):43--54, 2002.

\bibitem{TD11}
D~Tautz and T~Domazet-Lo{\v{s}}o.
\newblock The evolutionary origin of orphan genes.
\newblock {\em Nature Reviews Genetics}, 12(10):692--702, 2011.

\bibitem{TR04}
J~S Taylor and J~Raes.
\newblock Duplication and divergence: the evolution of new genes and old ideas.
\newblock {\em Annu. Rev. Genet.}, 38:615--643, 2004.

\bibitem{VF02}
A~V{\'a}zquez, A~Flammini, A~Maritan, and A~Vespignani.
\newblock Modeling of protein interaction networks.
\newblock {\em Complexus}, 1(1):38--44, 2002.

\bibitem{W03}
Andreas Wagner.
\newblock How the global structure of protein interaction networks evolves.
\newblock {\em Proceedings of the Royal Society of London B: Biological
  Sciences}, 270(1514):457--466, 2003.

\bibitem{W61}
HB~Wright.
\newblock {\em Tables if integrals and other mathematical data}.
\newblock The MacMillan Company, New York, 1961.

\end{thebibliography}

\vfill\eject

\end{document}